\documentclass{aa}
\usepackage{graphicx,amsmath,txfonts}

\begin{document}
\newcommand{\mearth}{M_\oplus}
\title{On the dynamics of resonant super-Earths in disks with turbulence driven by stochastic 
forcing}
\author{ A. Pierens \inst{1,2} \and C. Baruteau \inst{3,4} \and F. Hersant \inst{1,2} }
\institute{ Universit\'e de Bordeaux, Observatoire Aquitain des Sciences de l'Univers,
    BP89 33271 Floirac Cedex, France \label{inst1} \and
   Laboratoire d'Astrophysique de Bordeaux,
    BP89 33271 Floirac Cedex, France \label{inst2} \\
     \email{arnaud.pierens@obs.u-bordeaux1.fr}
\and Department of Astronomy and Astrophysics, University of Santa Cruz, CA 95064, United States \label{inst3}
  \and DAMTP, University of Cambridge, Wilberforce Road, Cambridge CB30WA, United Kingdom}
\abstract
{A number of sytems of multiple super-Earths have recently been discovered. Although the 
observed period ratios are generally far from strict commensurability, 
 the radio  pulsar PSRB$1257+12$ exhibits two near equal-mass planets of 
$\sim4$ $M_\oplus$ close to
being in a 3:2 mean motion resonance (MMR).  
}
{We investigate the evolution of a system of two super-Earths with masses 
$\le 4$ $M_\oplus$ embedded in a turbulent protoplanetary disk. The aim is to 
examine whether or not resonant trapping can occur  and be maintained in presence of turbulence and 
how this depends on the amplitude of the stochastic density fluctuations in the disk.}
{We have performed 2D numerical simulations using a grid-based hydrodynamical code in 
which turbulence is modelled as stochastic forcing. We assume that the outermost planet is initially located 
just outside the 
position of the 3:2 mean motion resonance  with the inner one and we study the dependance of the
resonance stability with the amplitude of the stochastic forcing. } 
{For systems of two equal-mass planets we find that in disk models  
with an effective viscous stress parameter $\alpha \sim 10^{-3}$, damping effects 
due to type I migration can counteract the effects of diffusion of the resonant angles, in 
such a way that the 3:2 resonance can possibly remain stable over the disk lifetime. For systems 
of super-Earths with mass ratio $q=m_i/m_o\le 1/2$, where $m_i$($m_o$) is the mass of the innermost 
(outermost) planet, the 3:2 resonance is broken in turbulent disks with effective viscous stresses
$2\times 10^{-4}\lesssim \alpha \lesssim 1\times 10^{-3}$ but the planets become 
locked in stronger $p+1$:$p$ resonances, with $p$ increasing as the value for $\alpha$ increases. 
For $\alpha \gtrsim 2\times 10^{-3}$, the evolution
can eventually involve temporary capture in a  8:7 commensurability but no stable MMR is formed.  }
{Our results suggest that for values of the viscous stress parameter typical to those generated by MHD 
turbulence, MMRs between two super-Earths are likely to be disrupted by stochastic density fluctuations. 
For lower levels of turbulence however, 
as is the case in presence of a dead-zone, 
resonant trapping can be maintained in systems with moderate values of the planet mass ratio.}
\maketitle
\section{Introduction}

To date, about $25$ extrasolar planets with masses less than $10\;M_\oplus$ and commonly 
referred  to as super-Earths have been discovered (e.g. {\tt http://exoplanet.eu}). 
Although two of them, Corot-7b (L\'eger et al. 
2009; Queloz et al. 2009) and GJ 1214b (Charbonneau et al. 2009) were detected via the transit
method, most of them were found by high-precision radial velocity surveys. It is expected that 
the number of observed super-Earths will considerably increase in the near future with the advent
of space observatories Corot and Kepler. \\
 Interestingly, Kepler team has recently announced the 
discovery of $\sim 170 $ multi-planet systems candidates (Lissauer et al. 2011), although 
these need to be confirmed by follow-up programs. 
Previous to Kepler results, four  multi-planet systems containing at least two super-Earths  
had been detected around PSR B1257+12, HD 69830, GJ 581 and HD 40307. For the sytems 
 around main-sequence stars (HD 69830, GJ 581, HD 40307), 
the observed period ratios between two adjacent low-mass planets are quite far from 
strict commensurability.  However, the planetary system that is orbiting the radio pulsar PSR B1257+12   
  exhibits two planets with masses $3.9$ $ M_\oplus$  and $4.3$ $M_\oplus$ in a 3:2 mean motion resonance 
(Konacki \& Wolszczan 2003). Papaloizou \&
Szuszkiewicz (2005) showed that, for this system, the existence of such a 
resonance can be understood by a model in which two low-mass planets with mass ratio close
to unity undergo convergent type I migration (e.g. Ward 1997; Tanaka et al. 2002) while still 
embedded in a gaseous laminar disk until capture in that resonance occurs. More generally, these 
authors found that, for more disparate mass ratios and provided that convergent migration occurs, 
the evolution of a system of two  planets
in the $1-4$ $M_\oplus$ mass range is likely to result in the formation of high first-order 
commensurabilities $p+1$:$p$ with $p\ge 3$. Studies aimed at examining the interaction of many 
embryos within protoplanetary disks also suggest that capture in resonance between adjacent cores
through type I migration appears as a natural outcome of such a system (McNeil et al. 2005; 
Cresswell \& Nelson 2006). This, combined with the fact that the majority of super-Earths 
are found in multiplanetary systems (Mayor et al. 2009), would suggest that systems of
resonant super-Earths are common. The fact that most of the multiple systems of super-Earths 
observed so far do not exhibit mean motion resonances may be explained by a scenario in 
which strict commensurability is lost due to circularization through tidal interaction with the central 
star as the planets migrate inward and pass through the disk inner edge (Terquem \& Papaloizou 2007).\\

 Moreover, it is expected that in presence of strong disk turbulence, effects arising 
from stochastic density flucuations will 
prevent super-Earths from staying 
in a resonant configuration. It is indeed now widely accepted that a source of anomalous viscosity
   due to turbulence is required to account for the estimated accretion rates for  Class II T Tauri 
stars, which are typically $\sim 10^{-8}$ M$_{\odot}${yr}$^{-1}$ 
(Sicilia-Aguilar et al. 2004). The origin of turbulence is believed to be related to 
the magneto-rotational instability (MRI, Balbus \& Hawley 1991) for which a number 
of studies (Hawley et al. 1996; Brandenburg et al. 1996)  have shown that the 
non-linear outcome of this instability is MHD 
turbulence with an effective viscous stress parameter $\alpha$ 
ranging between $\sim 5\times 10^{-3}$ and $\sim 0.1$, depending on the magnetic field amplitude and topology. \\
So far, the effects of stochastic density fluctuations in the disk on the evolution of  two-planet 
systems has received little attention. Rein 
\& Papaloizou (2009) developed an analytical model and performed N-body simulations of  two-planet systems subject to external 
stochastic forcing and showed that turbulence can produce systems in mean motion resonance with
broken apsidal corotation, explaining thereby the resonant configuration of the HD $128311$ 
system. Adams et al. (2008) examined the effets of turbulent torques on the survival 
of resonances using a pendulum model with an additional stochastic forcing term.  
They found that mean motion resonances are generally disrupted by 
turbulence within disk lifetimes. Lecoanet et al. (2009) extended this latter work 
 by considering disk-induced damping effets and planet-planet interactions. They found that 
systems with sufficiently large damping can maintain resonances and suggested that two-planet
systems composed of a Jovian outer planet plus a smaller inner planet are likely to remain bound in resonance. \\

In this paper we present the results of hydrodynamical simulations of systems  composed 
of two planets in the $1-4$ $M_\oplus$ mass 
range embedded in a protoplanetary disk in which turbulence is driven by stochastic forcing. Planets 
undergo convergent migration 
as a result of the underlying type I migration and we consider a scenario in which the initial 
separation between the planets is slightly larger than that 
corresponding to the 3:2 resonance. The aim of this work is to investigate whether 
or not resonant trapping can occur  and be maintained in turbulent disks and how the stability of the 
3:2 resonance depends on the amplitude of the turbulence-induced density fluctuations. 
We find that for systems of equal-mass planets the 3:2 resonance can be maintained 
provided that the level of turbulence is relatively weak, corresponding to a value for 
the effective viscous stress parameter of $\alpha\lesssim 10^{-3}$. In models  with mass ratios 
$q=m_/m_o\le 1/2$ however, where $m_i$ ($m_o$) is the mass of the inner (outer) planet, the 3:2 resonance 
is disrupted in presence of weak turbulence but the 
planets can become eventually locked in higher first-order commensurabilities. For a level of 
turbulence corresponding to $\alpha \sim 5 \times 10^{-3}$ however, MMRs are likely to be 
disrupted by stochastic density fluctuations. \\

This paper is organized as follows. In Sect. 2, we describe the
hydrodynamical model and the
numerical setup. In Sect.3, we use a simple model to  estimate the critical level of 
turbulence above which the 3:2 resonance would be unstable. In  Sect. 4 we present the results of our simulations.
We finally summarize and draw our conclusions in Sect. 5.

\section{The hydrodynamical model}

\subsection{Numerical method}
\label{sec:num}

In this paper, we adopt a 2D disk model for which all the physical 
quantities are vertically averaged. We work in a non-rotating frame,
and adopt cylindrical polar coordinates $(R,\phi)$ with the origin
located at the position of the central star. Indirect terms resulting
from the fact that this frame is non-inertial are incorporated in the
equations governing the disk evolution (Nelson et al. 2000). 
 Simulations were performed with the FARGO and GENESIS numerical codes. 
Both codes employ an advection scheme based on the monotonic transport algorithm 
(Van Leer 1977) and include the FARGO algorithm (Masset 2000) to avoid timestep limitation 
due to the Keplerian orbital velocity at the inner edge of the grid. The evolution of each
 planetary orbit is computed using a fifth-order Runge-Kutta integrator
 (Press et al. 1992) and by calculating the torques exerted by the disk
on each planet. We note that a softening parameter $b=0.6 H$, where $H$ is the disk scale height, 
is employed when calculating the planet potentials.\\

The computational units  that we adopt are such that the unit of mass is the 
central mass $M_*$, the unit of distance is the initial semi-major axis
$a_i$  of
the innermost planet and the unit of time is $(GM_*/a_i^3)^{-1/2}$.
In the simulations presented here, we use $N_R=256$ radial grid cells
uniformly distributed between $R_{in}=0.4$ and $R_{out}=2.5$ and 
$N_{\phi}=768$ azimuthal grid cells. Wave-killing zones are employed
 for $R<0.5$ and $R>2.1$ in order to avoid wave reflections at the disk edges (de Val-Borro et al. 2006).\\

In this work, turbulence is modelled by applying at each time-step a turbulent 
potential $\Phi_{turb}$ to the disk (Laughlin \& al. 
2004, Baruteau \& Lin 2010) and corresponding to the superposition of $50$ 
wave-like modes. This reads:

\begin{equation}
\Phi_{turb}(R,\phi,t)=\gamma R^2 \Omega^2\sum_{k=1}^{50}\Lambda_k(R,\phi,t),
\label{eq:phi}
\end{equation} 

with:

\begin{equation}
\Lambda_k=\xi_k e^{-\frac{(R-R_k)^2}{\sigma_k^2}}
\cos(m_k\phi-\phi_k-\Omega_k\tilde{t_k})
\sin(\pi \tilde{t_k}/\Delta t_k).
\label{eq:lambda}
\end{equation}

In Eq. \ref{eq:lambda}, $\xi_k$ is a dimensionless constant  parameter randomly 
sorted with a Gaussian distribution of unit width. 
$R_k$ and $\phi_k$ are, respectively, the radial and
azimuthal initial coordinates of the mode with wavenumber $m_k$,
 $\sigma_k=\pi R_k /4m_k$ is the radial extent of that mode and
$\Omega_k$ denotes the keplerian angular velocity at $R=R_k$.  
$R_k$ and $\phi_k$ are both randomly sorted  with a uniform 
distribution whereas $m_k$ is randomly sorted with a logarithmic
distribution between $m_k=1$ and $m_k=96$. Following Ogihara et al. (2007), 
 we set $\Lambda_k=0$ if $m_k > 6$ in order to save computing time.
Each mode of wavenumber $m_k$ starts at time $t=t_{0,k}$ and 
 terminates when $\tilde{t_k}=t-t_{0,k} > \Delta t_k $,
where $\Delta t_k=0.2\pi R_k /m_k c_s$, $c_s$ being the sound speed, 
denotes the lifetime of mode with wavenumber $m_k$.  Such a value for 
$\Delta t_k$ yields a turbulence with autocorrelation timescale $\tau_c\sim T_{orb}$, 
where $T_{orb}$ is the orbital period at $R=1$ (Baruteau \& Lin 2010). 
  \\ 
In Eq. \ref{eq:phi}, $\gamma$ denotes the value of the turbulent
forcing parameter which controls the amplitude of the stochastic density perturbations.
In the simulations presented here, we used four different values for 
$\gamma$ namely: $\gamma=6\times 10^{-5}$, $1.3\times 10^{-4}$, $1.9\times 10^{-4}$,  
$3\times 10^{-4}$. Given that $\gamma$ is related to 
the effective viscous stress parameter $\alpha$ and the disk aspect ratio 
 $h=H/R$ by the relation $\alpha=1.4\times 10^2(\gamma/h)^2$ 
(Baruteau \& Lin 2010), the latter values for $\gamma$ correspond to  
$\alpha \cong 2\times 10^{-4}$, $10^{-3}$, $2\times 10^{-3}$, $5\times 10^{-3}$ 
respectively. Inviscid simulations with $\alpha=0$ were
also performed for comparison. \\

 In calculations with high values of $\gamma$, viscous stresses arising from turbulence  
can eventually lead to a significant change in the disk surface density profile over 
a few thousand orbits. This is also observed in three dimensional MHD 
simulations in which turbulence is generated by the MRI (Papaloizou \& Nelson 2003). For lower 
values of $\gamma$, such an effect also occurs but over a much
longer timescale.
In order to examine how this affects the results of the simulations, we have performed 
additional simulations in which the initial surface density profile is restored on a 
characteristic timescale $\tau_m$. We follow Nelson \& Gressel (2010) and solve 
the following equation at each timestep:
\begin{equation}
\frac{\partial \Sigma}{\partial t}=-\frac{\Sigma-\Sigma_{init}}{\tau_m}
\label{eq:damp}
\end{equation}
where $\Sigma_{init}$ is the initial disk surface density and where $\tau_m$ was set to $\tau_m=20$ orbits. 
Such a value is shorter than the viscous timescale but longer than 
both the dynamical timescale at the outer edge of the disk and the lifetime of the 
mode with wavenumber $m=1$.  The results of such simulations are discussed in Sect. \ref{sec:comparison}.

\subsection{Initial conditions}

In this paper, we adopt a locally isothermal equation of state with a
 fixed temperature profile given by 
$T=T_0R^{-\beta}$ where $\beta=1$ and where $T_0$ is the temperature at $R=1$. 
This corresponds to a disk with constant aspect ratio $h$ and for most of the 
simulations, we choose $T_0$ so that  
$h=0.05$.  The initial surface density
profile is chosen to be $\Sigma_{init}(R)=\Sigma_0R^{-\sigma}$ with $\sigma=0.5$
 and we have performed
simulations with $\Sigma_0=2\times 10^{-4}$ and $\Sigma_0=4\times 10^{-4}$.
Assuming that the radius $R=1$ in the computational domain correponds to 
$5.2$ AU, such values for $\Sigma_0$ correspond to disks containing 
$0.02$ $M_\star$ and $0.04$ $M_\star$ respectively of gas material interior 
to $40$ AU. No kinematic viscosity is employed in all the runs presented
here.\\    

\begin{table}
\centering{}
\begin{tabular} {c c c c c}
\hline \hline
Model & $m_i\;(\mearth)$ &  $m_o\; (\mearth)$ & $\Sigma_0$ & $h$ \\
\hline
$G1$ & $3.3$ & $3.3$ & $2\times 10^{-4}$ &  $0.05$ \\
$G2$ & $3.3$ & $3.3$  & $4\times 10^{-4}$ & $0.05$\\
$G3$ & $3.3$ & $3.3$  & $2\times 10^{-4}$ & $0.04$\\
$G4$ & $1.6$ & $1.6$  & $2\times 10^{-4}$ &  $0.05$\\
$G5$ & $1.6$ & $3.3$  & $2\times 10^{-4}$ & $0.05$\\
\hline
\end{tabular}
\caption{Parameters used in the simulations}
\label{table1}
\end{table}

The inner and outer planets initially evolve on circular orbits  at $a_i=1$ and $a_o=1.33$ respectively, 
which corresponds to a configuration for which 
the outermost planet is initially located just outside the 3:2  MMR with the inner one. 
For most models,
we focus on  equal-mass planets with $m_i=m_o\le 3.3$ $M_\oplus$, where $m_i$ ($m_o$) is the mass of 
innermost (outermost) planet. However, we have also considered one case in which the planet mass 
ratio $q=m_i/m_o$ is $q=1/2$. 
The parameters for all models we conducted are  summarized in Table \ref{table1}.
Given that the type I 
migration timescale $\tau_{mig,p}$ of a planet with mass $m_p$, semimajor axis $a_p$  and on
a circular orbit with angular frequency $\Omega_p$
can be estimated in the locally isothermal limit by (Paardekooper et al. 2010):
\begin{equation}
\tau_{mig,p}=(1.6+\beta+0.7\sigma)^{-1} \frac{M_\star}{m_p}\frac{M_\star}{\Sigma(a_p) r_p^2}h^2\Omega_p^{-1},
\label{eq:taumig}
\end{equation}
we expect that equal low-mass planets embedded in our disk model will undergo convergent 
migration and become eventually trapped in the 3:2 resonance. For a larger 
initial separation between the two planets, capture in 2:1 resonance may also occur. 
 However, test simulations have shown that unless the disk mass is very low, differential migration is 
not slow enough for the planets 
to become trapped in that resonance.
This justifies our assumption that the planets are initially located just outside the 
3:2 resonance. We also comment that equal-mass planets migrating in the type I regime 
will undergo convergent migration provided that $\sigma < 3/2$ whereas $\sigma > 3/2$ will 
lead to divergent migration.

\section{Theoretical expectations} 
\begin{figure}
\centering
\includegraphics[width=0.9\columnwidth]{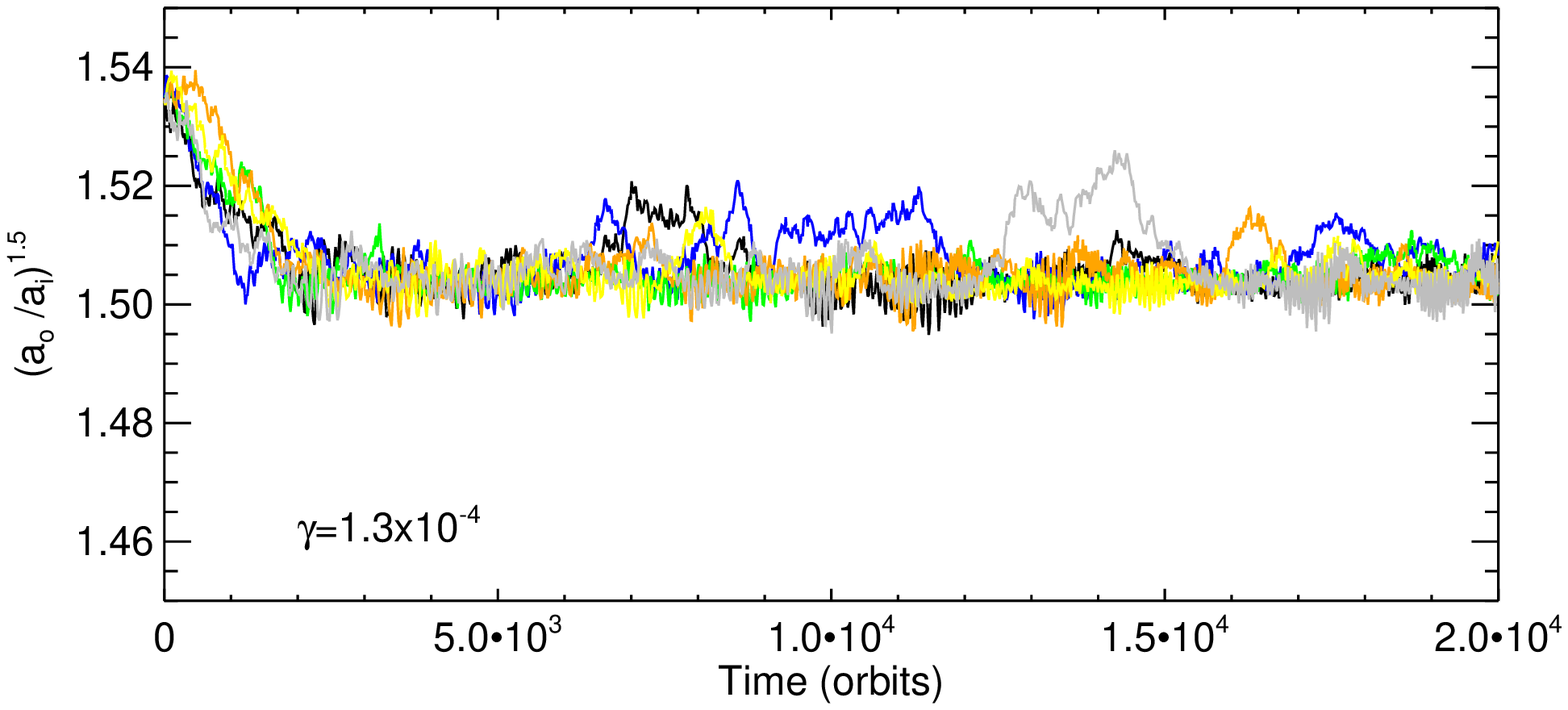}
\includegraphics[width=0.9\columnwidth]{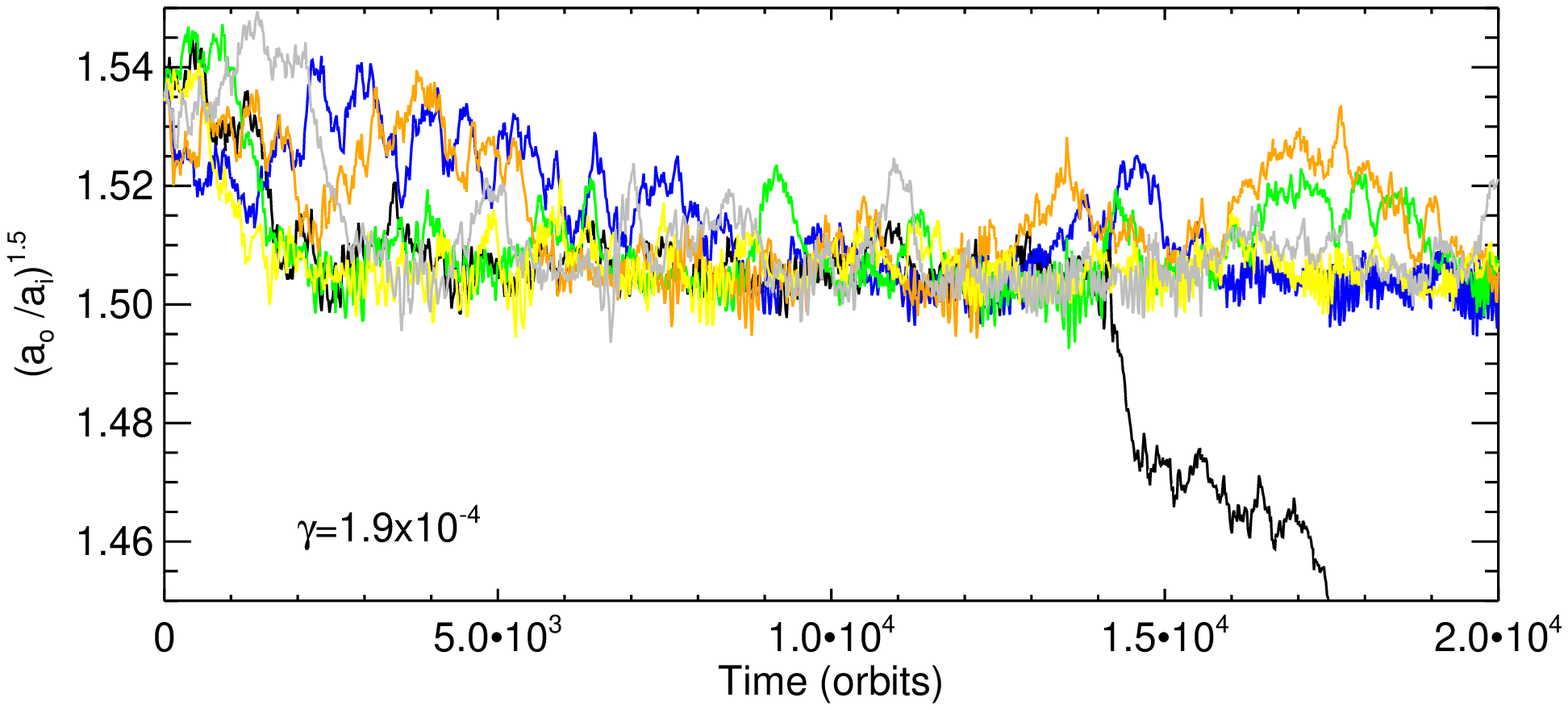}
\includegraphics[width=0.9\columnwidth]{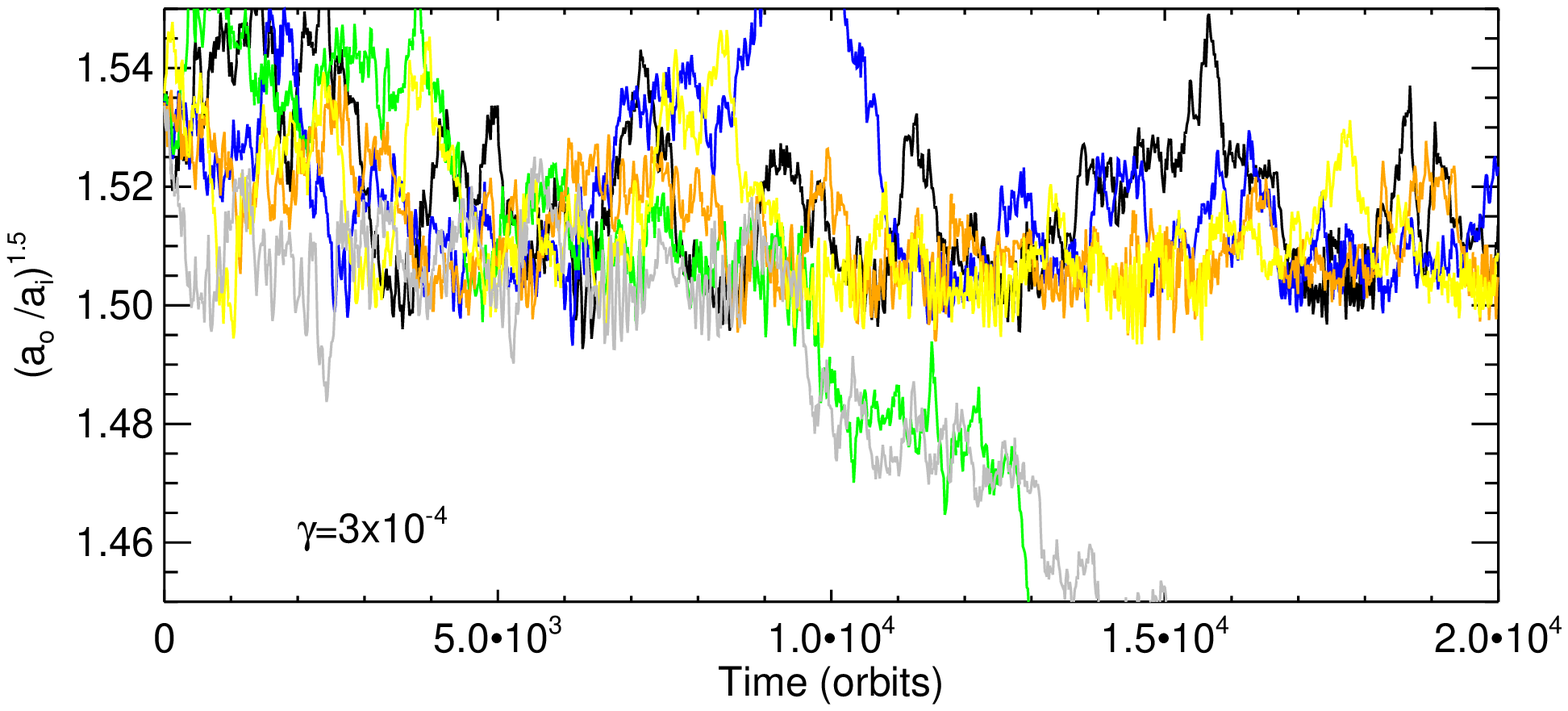}
\caption{Time evolution of the period ratio resulting from N-body runs 
for model G1 and for six 
different realizations with $\gamma=1.3\times 10^{-4}$, $\gamma=1.9\times 10^{-4}$, and
$\gamma=3\times 10^{-4}$. }
\label{nbody}
\end{figure}
\begin{figure}
\centering
\includegraphics[width=0.49\columnwidth]{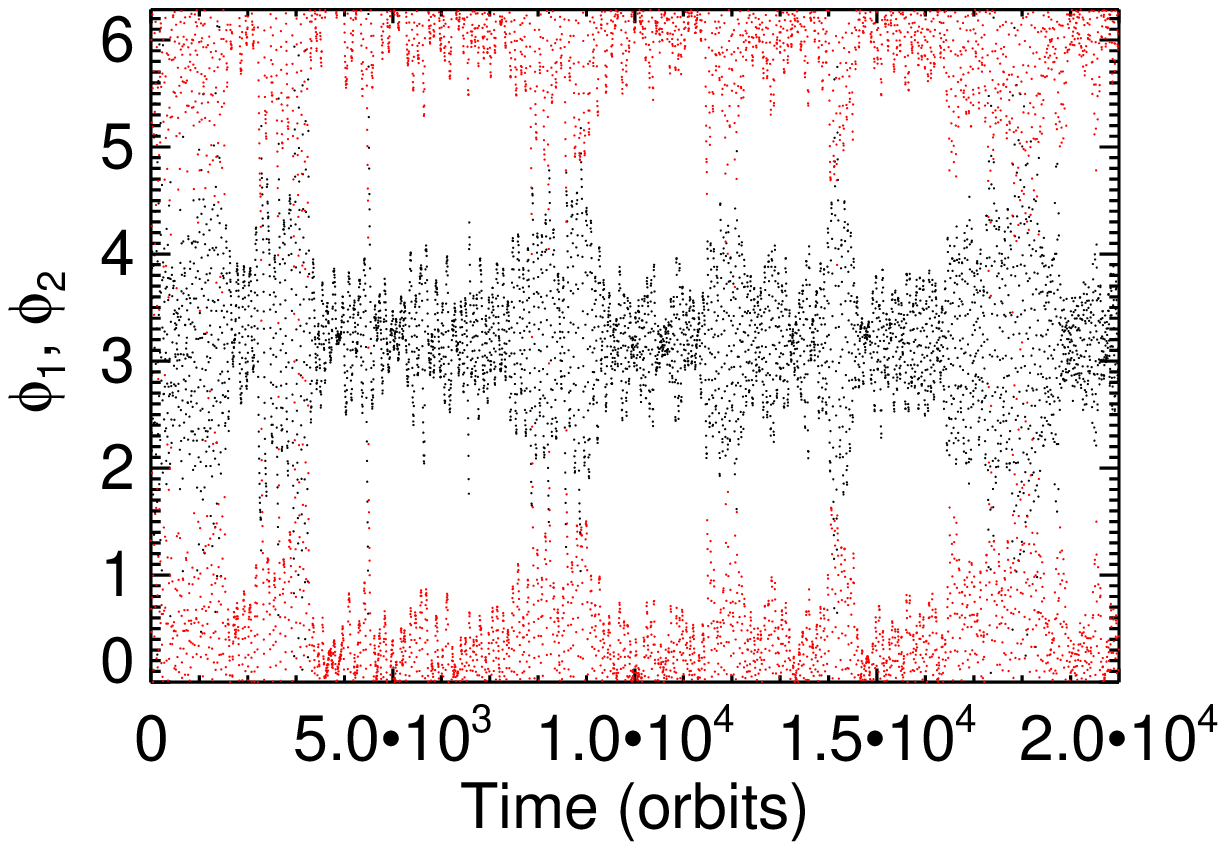}
\includegraphics[width=0.49\columnwidth]{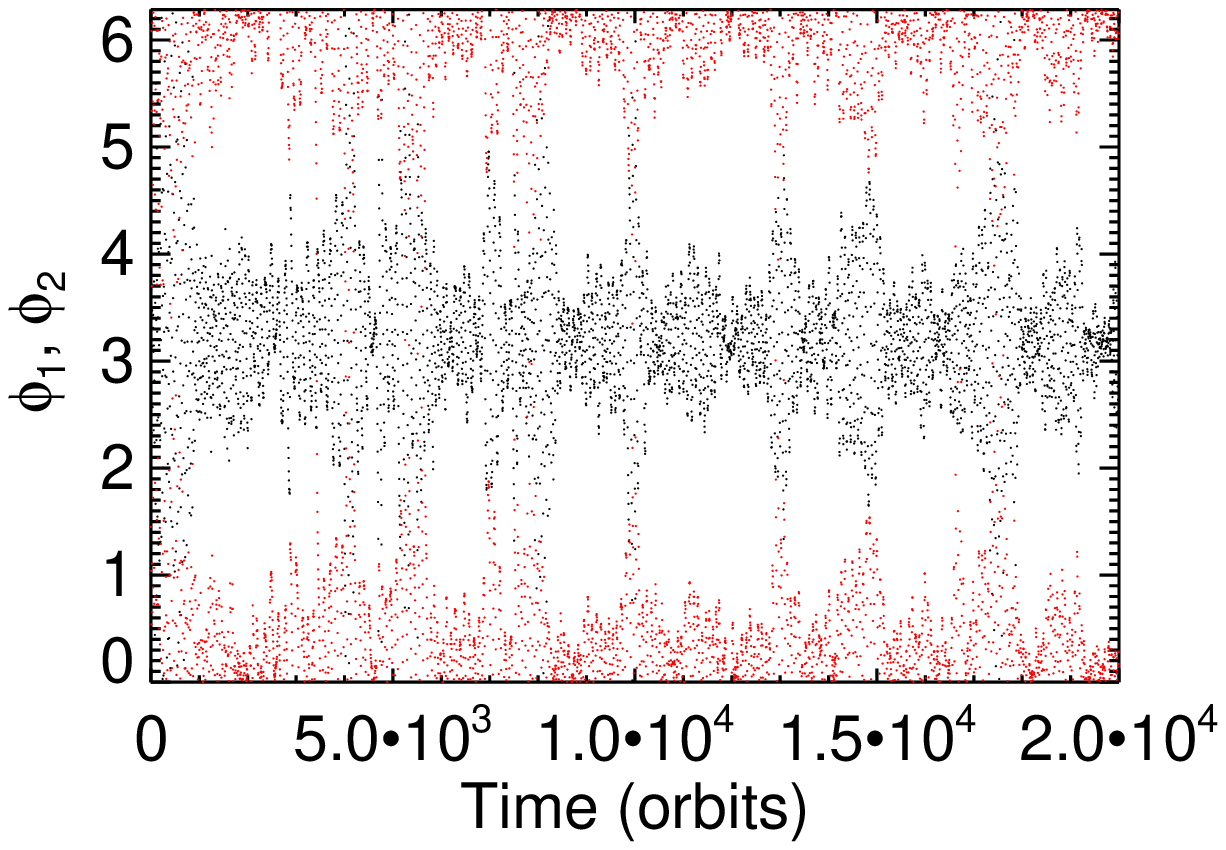}
\includegraphics[width=0.49\columnwidth]{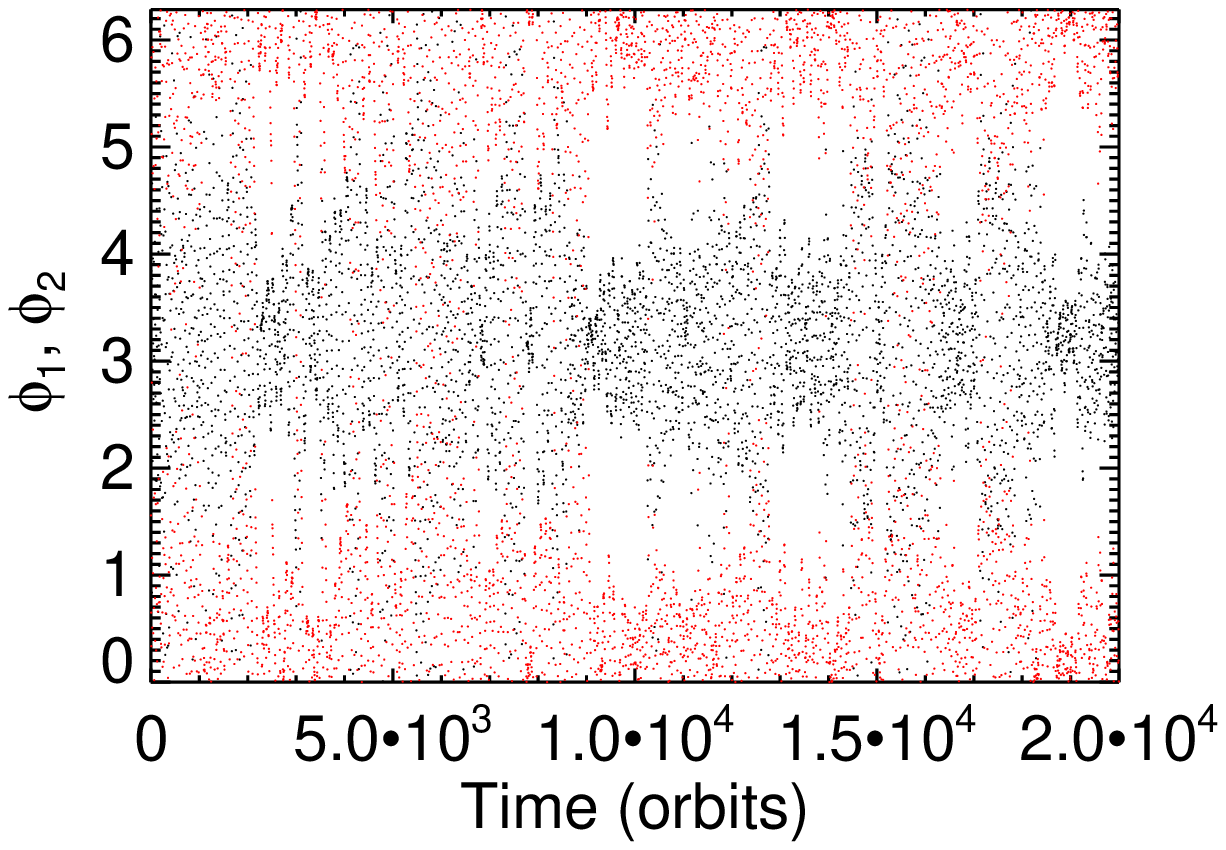}
\includegraphics[width=0.49\columnwidth]{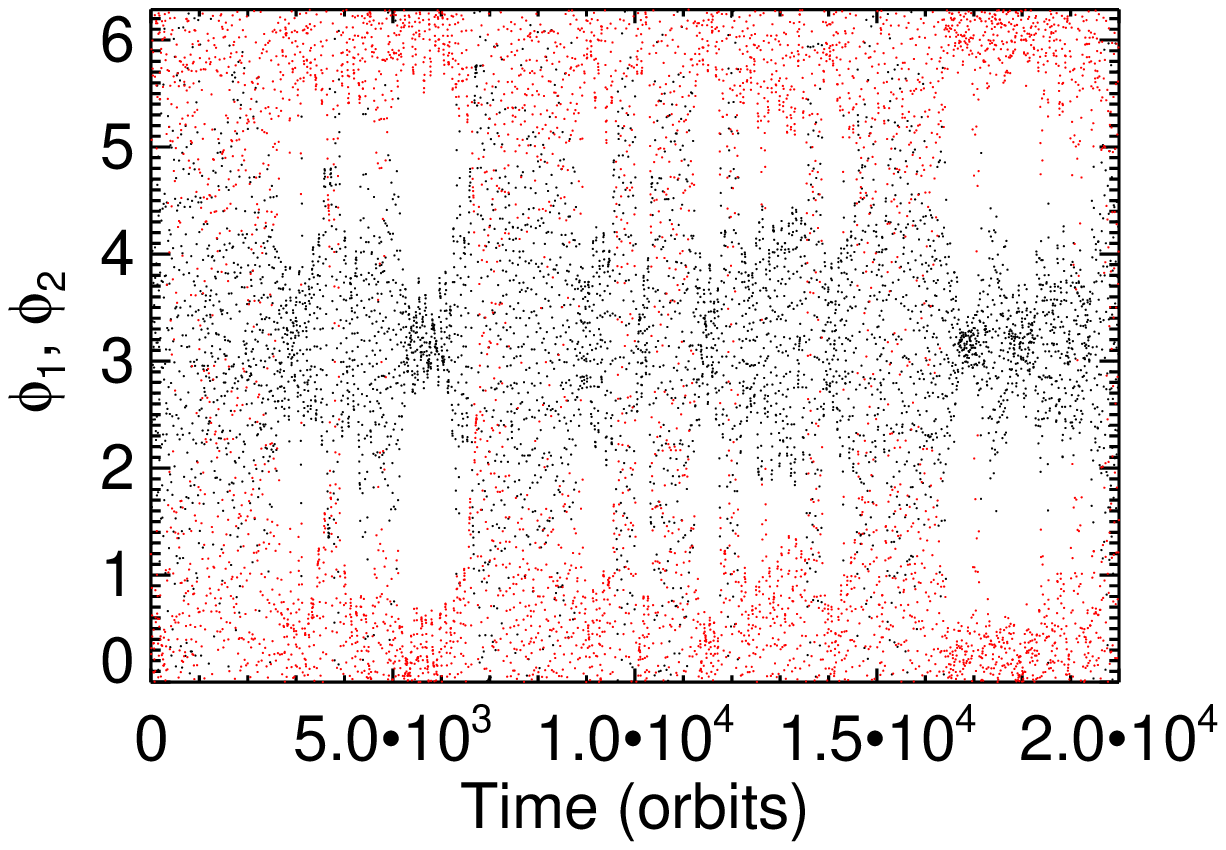}
\includegraphics[width=0.49\columnwidth]{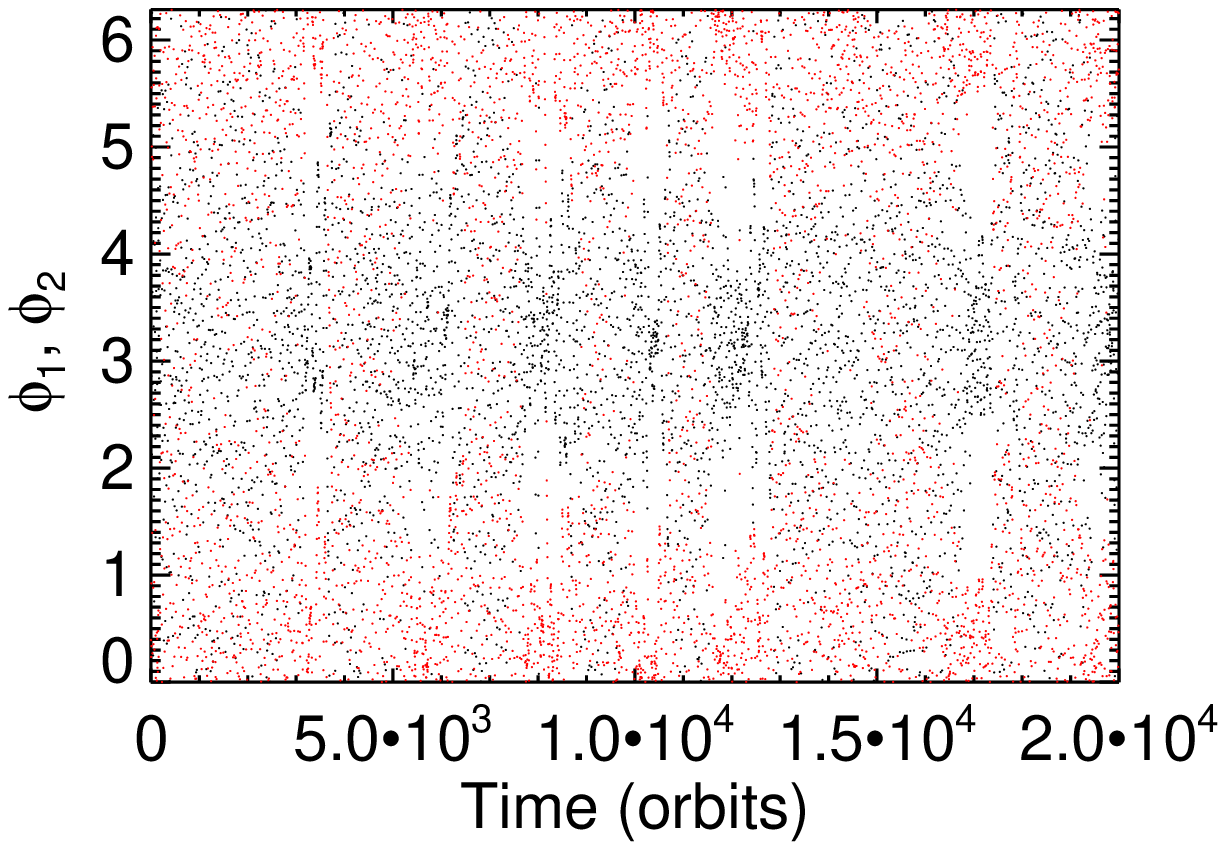}
\includegraphics[width=0.49\columnwidth]{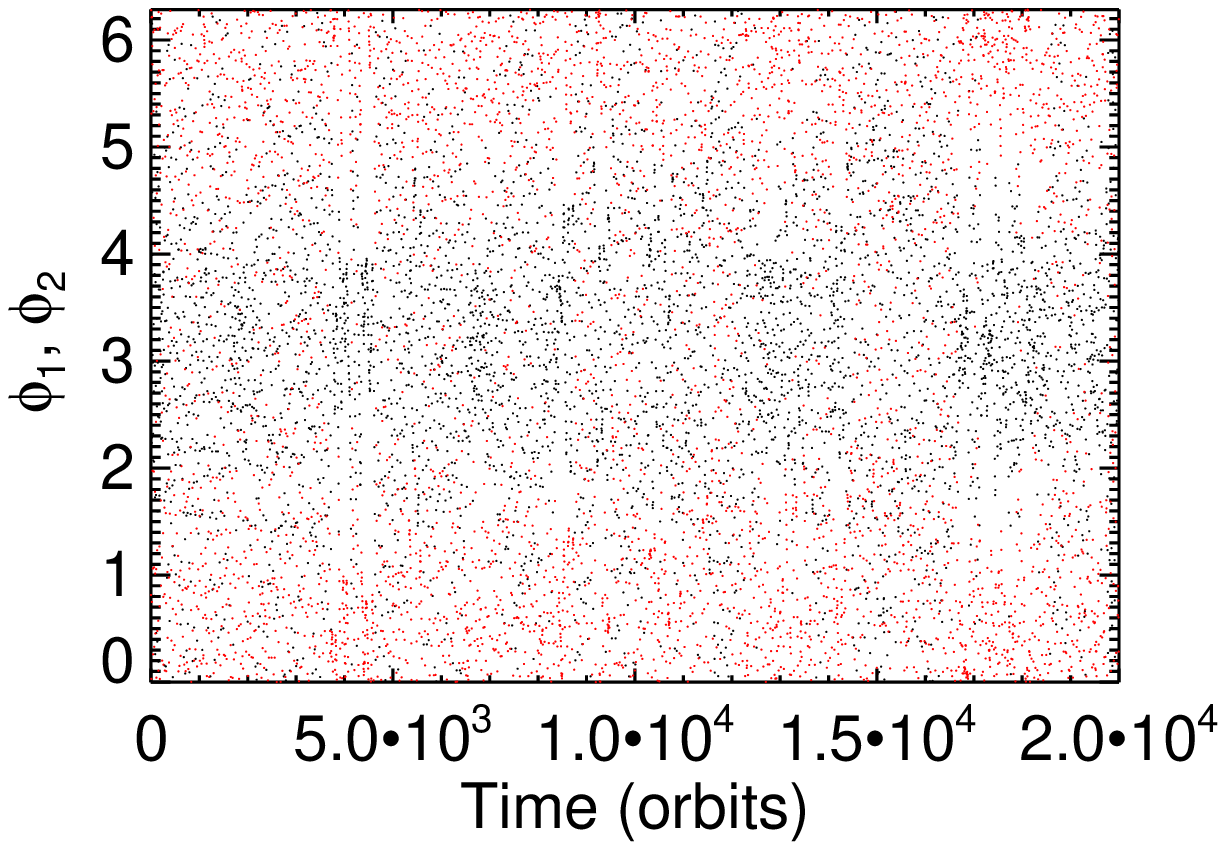}
\caption{{\it Upper panel:} time evolution of the 
resonant angles $\phi_1=3\lambda_o-2\lambda_i-\omega_i$ (black) and 
$\phi_2=3\lambda_o-2\lambda_i-\omega_o$ (red) resulting from N-body runs for model G1 and for two different 
realizations with $\gamma=1.3\times 10^{-4}$. {\it Middle panel:} same but for $\gamma=1.9\times 10^{-4}$. 
{\it Lower panel:} same but for $\gamma=3\times 10^{-4}$. }  
\label{nbody-angles}
\end{figure}
\label{sec:exp}
In this section, we consider two low-mass planets embedded in a turbulent disk
 and locked in a $p+1$:$p$ mean motion resonance, and we derive the critical amplitude of the turbulent forcing 
$\gamma_c$ below which the resonance would remain stable.  Following Adams et al. (2008) 
 and Rein \& Papaloizou (2009),  
we assume that only the outermost planet experiences the torques arising from the disk. 
We also assume that the planets have near equal mass, 
in order to avoid the chaotic regime which comes into play for disparate masses (Papaloizou \&
Szuszkiewicz 2005). In the limit of large damping
rate for the resonance and neglecting effects from planet-planet interaction, the 
asymptotic value $P$ of the probability for the resonance to be maintained is given by (Lecoanet et al. 2009):

\begin{equation}
P=4\left(\frac{1}{\pi\tau_{d} D_\phi}\right)^{1/2},
\end{equation}

where  $D_\phi$ is  the diffusion coefficient associated with the
resonant angle diffusion  and $\tau_d$ is the damping timescale for the resonance angle. 
 This equation is valid at late times $t\gg \omega_0^{-1}$ where $\omega_0$ is the libration frequency 
of the resonant angles, as long as $t \gg D_\phi^{-1}$ and $D_\phi^{-1}\gg \tau_d$. From the previous equation, we can estimate the maximum value of the diffusion
coefficient for the sytem to remain bound in resonance with probability $P=1$. This reads:
\begin{equation}
D_\phi=\frac{16}{\pi}\tau_d^{-1}.
\label{eq:dphicrit}
\end{equation}
As shown in Adams et al. (2008),  $D_\phi$ can be related to the diffusion coefficient $D_{H,o}$
associated with the diffusion of the outer planet's angular momentum as:
\begin{equation}
D_\phi=\frac{D_{H,o}}{9m_o^2\omega_0^2a_o^4}.
\label{eq:dphi}
\end{equation}
For moderate eccentricities, $\omega_0$ is given by:
\begin{equation}
\omega_0^2=-3j_2^2{\cal C}\Omega_i e_i^{|j_4|}\quad \text{with} \quad {\cal C}=q_0\Omega_i \alpha f_d(\alpha),
\label{eq:frequency}
\end{equation}
where  $e_i$ is the eccentricity of the inner planet, $q_0=m_0/M_\star$ and $\Omega_i$ is the angular frequency of the innermost planet. In the previous equation, 
$\alpha=a_i/a_o$, $(j_2,j_4)$ are integers which depend on the resonance being considered and 
$f_d(\alpha)$ results from the expansion of the disturbing function. In the case of the 
3:2 resonance, we have $j_2=-2$, $j_4=-1$ and $\alpha f_d(\alpha) \sim -1.54$  (Murray \& Dermott 1999). \\

In Eq. \ref{eq:dphi}, $D_{H,o}$ can be expressed in terms of both the correlation timescale $\tau_c$ associated with 
the stochastic torques exerted on the  outer planet and the standard deviation of the turbulent torque distribution $\sigma_{t}$ as:
\begin{equation} 
D_{H,o}=\sigma_t^2 \tau_c.
\label{eq:dj}
\end{equation}

 As discussed in Baruteau \& Lin (2010), $\sigma_t$ takes the following form when applied to the 
 outermost planet:

\begin{equation}
\sigma_t=C\Sigma_o q_o \gamma a_o^4 \Omega_o^2,
\label{eq:sigmat}
\end{equation}

where $q_o=m_o/M_\star$, $\Sigma_o$ is the value  of  the surface density at the position 
of the outer planet, $\Omega_o$ is the angular frequency of this planet and $C$ is a constant. 
For a simulation using the same disk parameters 
as for model G1 and with $\gamma=6\times 10^{-5}$,  we find $C\sim 1.6\times 10^2$, which is close to the
value
found by Baruteau \& Lin (2010). 
Combining Eqs. \ref{eq:dphi}, \ref{eq:dj} and \ref{eq:sigmat} gives an expression for the diffusion
coefficient associated with the diffusion of the resonant angle $D_\Phi$, in terms of the value 
for the turbulent forcing $\gamma$. This reads:
\begin{equation}
D_\phi=\frac{2 C^2 q_d^2 \gamma^2\Omega_o^3}{9\pi \omega_0^2},
\end{equation}
with $q_d=\pi \Sigma_o a_o^2/M_\star$. Setting $\delta \omega=\omega_0/ \Omega_o$, we can rewrite the previous expression as:
\begin{equation}
D_\phi=\frac{2C^2 q_d^2 \gamma^2\Omega_o}{9\pi \delta\omega^2}.
\end{equation}
We notice that in the case where $p=1$, $\delta \omega$ is comparable to the dimensionless width of the
libration zone.
Using the previous equation together with Eq. \ref{eq:dphicrit}, we find that the critical value
for the turbulent forcing above which the 3:2 resonance is disrupted is given by:
\begin{equation}
\gamma_c\sim 5.3\times 10^{-2}\frac{\delta\omega}{q_d}{(\tau_d \Omega_o)}^{-1/2}.
\end{equation}
In absence of turbulent forcing, we expect the amplitude of the resonant angles to scale as 
$\Omega_o^{-1/2}$ (Peale 1976).   According to Rein \& Papaloizou (2009), this implies that the damping 
timescale of the libration amplitude $\tau_d$ is twice the migration timescale $\tau_{mig}$ of the whole 
system,  namely that composed of the two planets locked in resonance and migrating inward together. 
In that case, the previous equation becomes:
\begin{equation}
\gamma_c\sim 3.7\times 10^{-2}\frac{\delta\omega}{q_d}{(\tau_{mig} \Omega_o)}^{-1/2}.
\label{eq:gammac}
\end{equation}

 In the limit where $m_i\sim 0$, we would have $\tau_{mig}=\tau_{mig,o}$ where 
$\tau_{mig,o}$ is the migration timescale of the outer planet which is given by Eq. \ref{eq:taumig} with 
$h=0.05$, $\sigma=0.5$ and $\beta=1$. 
In that case the 
expression for $\gamma_c$  becomes for our disk model:
\begin{equation}
\gamma_c \sim 3.5\times 10^{-2}\delta\omega q_o^{1/2}q_d^{-1/2}h^{-1}.
\label{eq:gammac2}
\end{equation}

 In order to check the validity of the previous expression for $\gamma_c$, we have performed a few 
 N-body runs using a fifth-order Runge Kutta method. In these calculations, the  
forces arising from type I migration are not determined self-consistently but  modelled 
using prescriptions for both the migration rate $\tau_{a_p}$ 
and eccentricity 
damping rate $\tau_{e_p}$ of the planets. For $\tau_{a_p}$, we used:
\begin{equation}
\tau_{a_p}=\left(\frac{1+(e_p/h)^5}{1-(e_p/h)^4}\right)\tau_{mig,p},
\end{equation} 
where $e_p$ is the planet eccentricity and where  $\tau_{mig,p}$ is given by Eq. \ref{eq:taumig} and where the numerical factor accounts for the modification of the migration rate at large eccentricities (Papaloizou \& Larwood 2000). For $\tau_{e_p}$ 
we used (Tanaka \& Ward 2004):
\begin{equation}
\tau_{e_p}=\frac{K h^2}{0.78}\left(1+0.25(e_p/h)^3\right) \tau_{mig,p}.
\end{equation}
In the last equation, $K\sim 1.7$ is a constant which was chosen in such a way  that the eccentricity damping
rate obtained in N-boby runs gives good agreement with that resulting from hydrodynamical simulations. Following 
Rein et al. (2010), we model effects of turbulence as an uncorrelated noise by  perturbing at each time step $\Delta t$ 
the velocity components $v_{i,p}$ of each planet by $\Delta v_{i,p}=\sqrt{2D\Delta t}\xi$ where 
$\xi$ is a random variable with gaussian distribution of unit width. $D$ is the diffusion coefficient 
which should vary as the planets migrate but which was fixed here to a constant value of 
$D=\sigma_t^2\tau_c/a_o^2$. \\
In Fig. \ref{nbody} we show the time
evolution of the orbital period ratio for N-body simulations with parameters corresponding to model G1 and
 for six different realizations with $\gamma=1.3\times10^{-4}, 1.9\times10^{-4}, 3\times10^{-4}$. 
 For this model, we estimate $\delta \omega\sim 1.9\times 10^{-3}$ (see Sect. \ref{sec:comparison}) which leads
to $\gamma_c\sim 2.5\times 10^{-4}$ using Eq. \ref{eq:gammac2}.   
From Fig. \ref{nbody}, it appears that capture in the 3:2 MMR occurs for most of the realizations with $\gamma \le 1.9\times 10^{-4}$. For two specific realizations of each value 
for $\gamma$ we considered,  the time evolution of the resonant angles  $\phi_1=3\lambda_o-2\lambda_i-\omega_i$ and
$\phi_2=3\lambda_o-2\lambda_i-\omega_o$ associated with the 3:2 resonance, where $\lambda_i$
($\lambda_o$) and $\omega_i$ ($\omega_o$) are respectively the mean longitude and
longitude of pericentre of the innermost (outermost) planet, is displayed in Fig. \ref{nbody-angles}. Although 
the angles can eventually circulate for short periods of time, it is clear that the 3:2 MMR remains stable 
on average for $\gamma \le 1.9\times 10^{-4}$.  For 
$\gamma=3\times 10^{-4}$, we find that the planets pass through the 3:2 resonance in two of the six 
realizations performed while the four other  can eventually involve temporarily capture in the 3:2 MMR.  In these cases however, 
the lifetime of the resonance does not exceed a few hundred orbits, as can be seen in the lower panel of Fig. \ref{nbody-angles} which 
displays for $\gamma=3\times 10^{-4}$ the time evolution of $\phi_1$ and $\phi_2$ for two realizations in which the period ratio 
 remains close to that corresponding to the 3:2 MMR. Therefore, the results of these N-body calculations 
suggest that the 3:2 MMR is only marginally  stable for such a value of $\gamma$, which is consistent  with the 
aforementioned analytical estimation of $\gamma_c\sim 2.5\times 10^{-4}$.

\begin{figure*}
\centering
\includegraphics[width=0.8\textwidth]{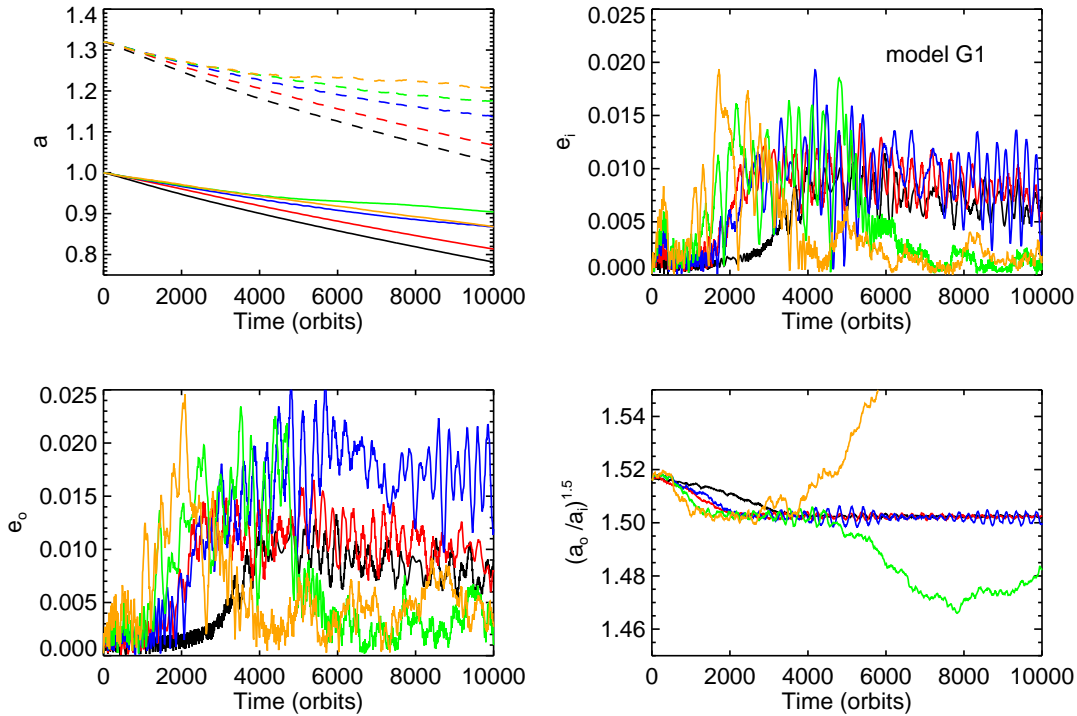}
\caption{{\it Upper left (first) panel}: time evolution of planet 
semi-major axes for model G1 and for the different values of $\gamma$ we considered namely for 
$\gamma=0$ (black line), $\gamma=6\times 10^{-5}$ (red line), $\gamma=1.3\times 10^{-4}$ (blue), 
$\gamma=1.9\times 10^{-4}$ (green) and $\gamma=3\times 10^{-4}$ (orange). {\it Upper right (second) 
panel}: time evolution of the inner planet eccentricity. {\it Third panel:} time evolution of the 
outer planet eccentricity. {\it Fourth panel:} time evolution of the period ratio. Simulations were
performed with GENESIS.}
\label{run1}
\end{figure*}

\section{Results of hydrodynamical simulations}

For equal mass planets ($q=1$),  results of hydrodynamical simulations suggest
that capture in 3:2  resonance can occur in  turbulent disks for which the level of 
turbulence is relatively weak. For systems with $q\le 1/2$ however, it appears that trapping in the 3:2 
resonance is maintained only provided that the disk is close  to being inviscid.

\subsection{Models with $q=1$}
For inviscid simulations with equal low-mass planets, the ability for the
two planets to become trapped in the 3:2 resonance depends mainly on the planets' relative migration rate
 which scales as $h^{-2}$. For model G3 ($h=0.04$), we find that capture in 3:2 resonance does not 
occur in that case due to the relative
migration timescale being shorter than the libration
period corresponding to that resonance. For other models  with $h=0.05$ however, 
it appears that the system can enter in a 3:2
commensurability which remains stable for the duration of the simulation, 
which generally covers $\sim 10^4$ orbits at $R=1$. This occurs not only 
for laminar disks, but also for turbulent disks provided that the value for the turbulent
forcing is not too large. For example, we find that the 3:2 commensurability is maintained in 
  most of the turbulent runs with $\gamma \le 1.3 \times 10^{-4}$ in models G1 and G2 whereas in model G4 
this occurs provided that $\gamma \le 6\times 10^{-5}$. Below we describe in more details the 
results of the simulations with $q=1$ and we use model G1 to illustrate how the evolution depends
on the value for the forcing parameter $\gamma$. 
\subsubsection{Orbital evolution}
\label{sec:orbit}
\begin{figure*}
\centering
\includegraphics[width=0.8\columnwidth]{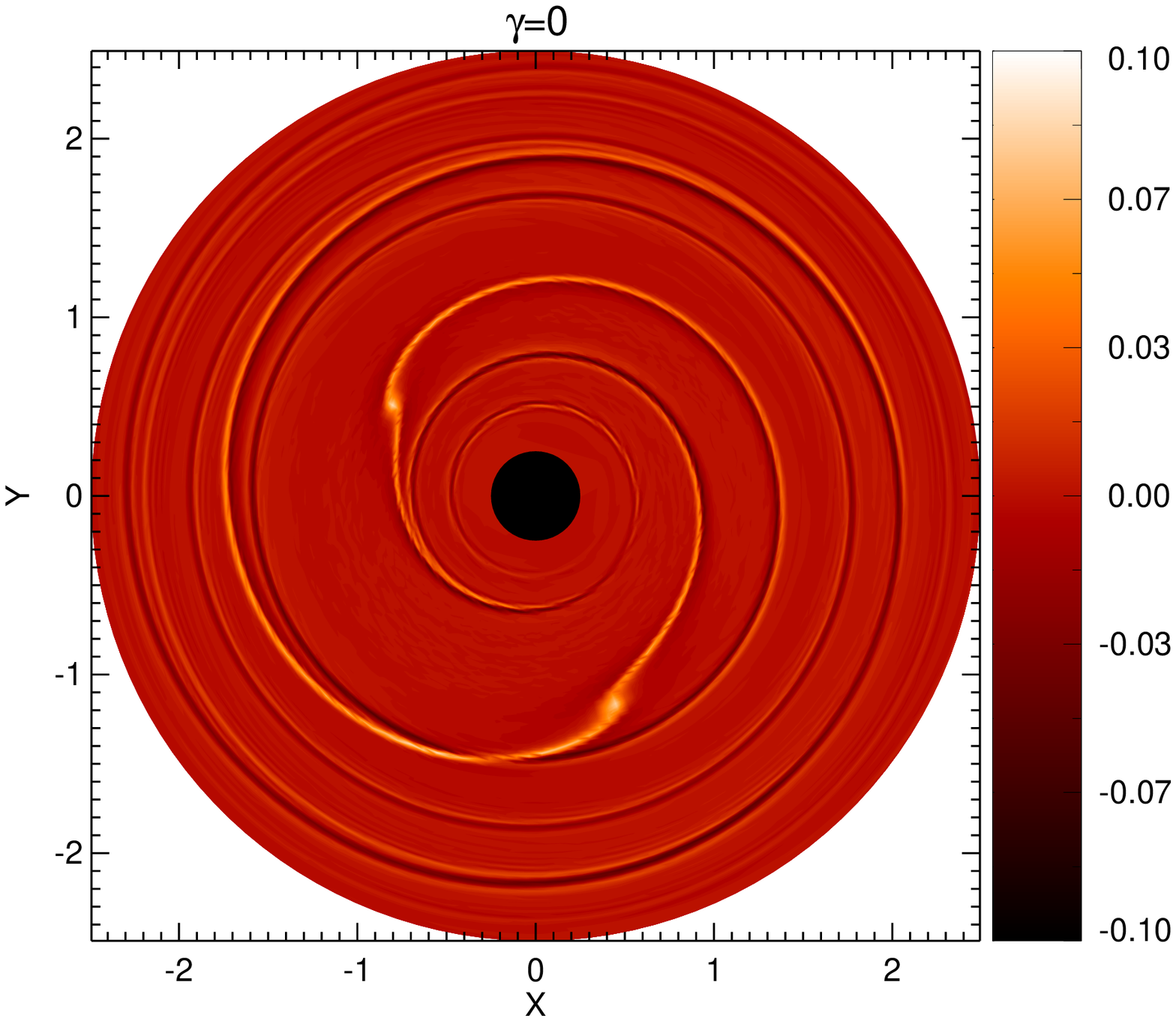}
\includegraphics[width=0.8\columnwidth]{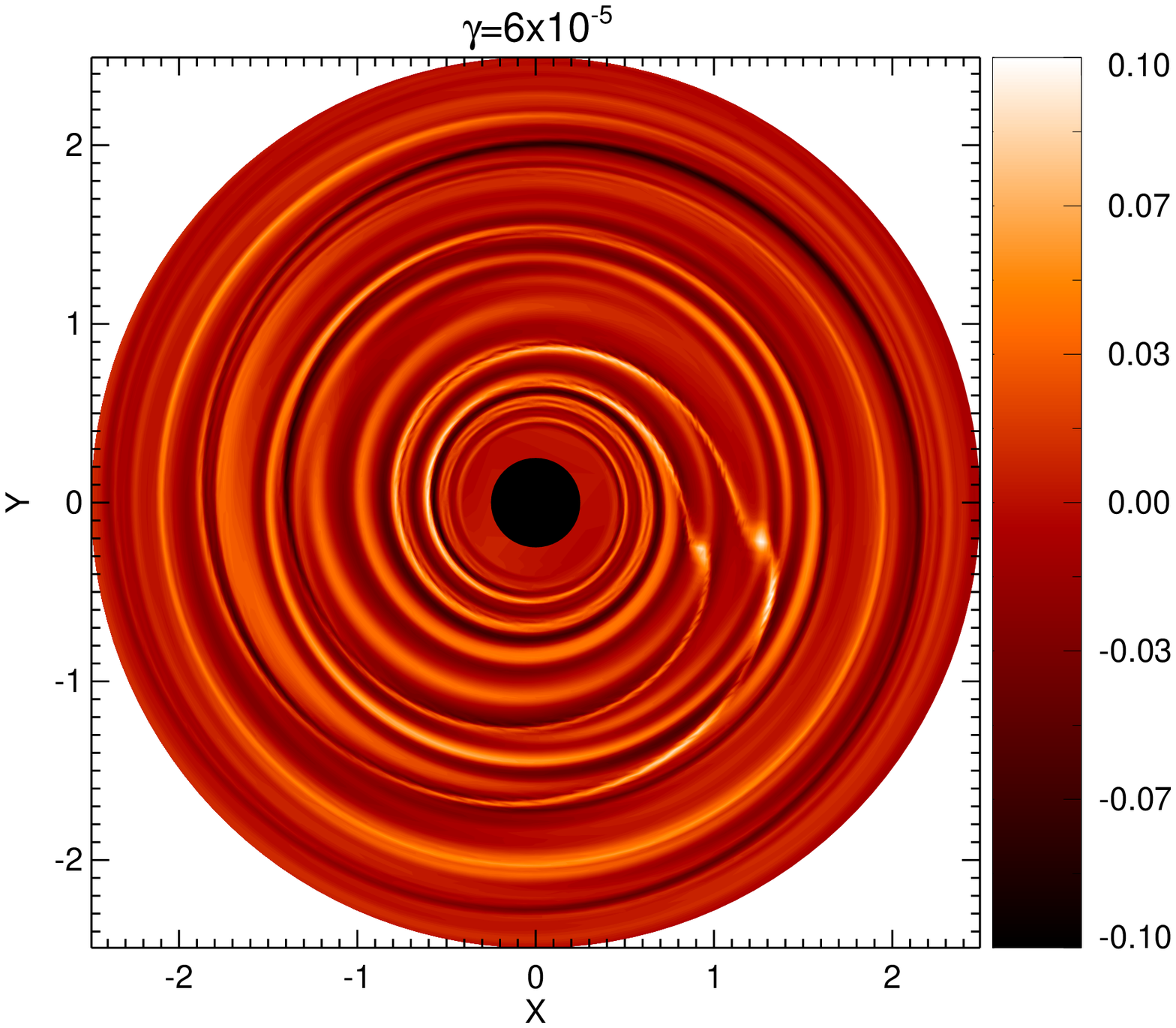}
\includegraphics[width=0.8\columnwidth]{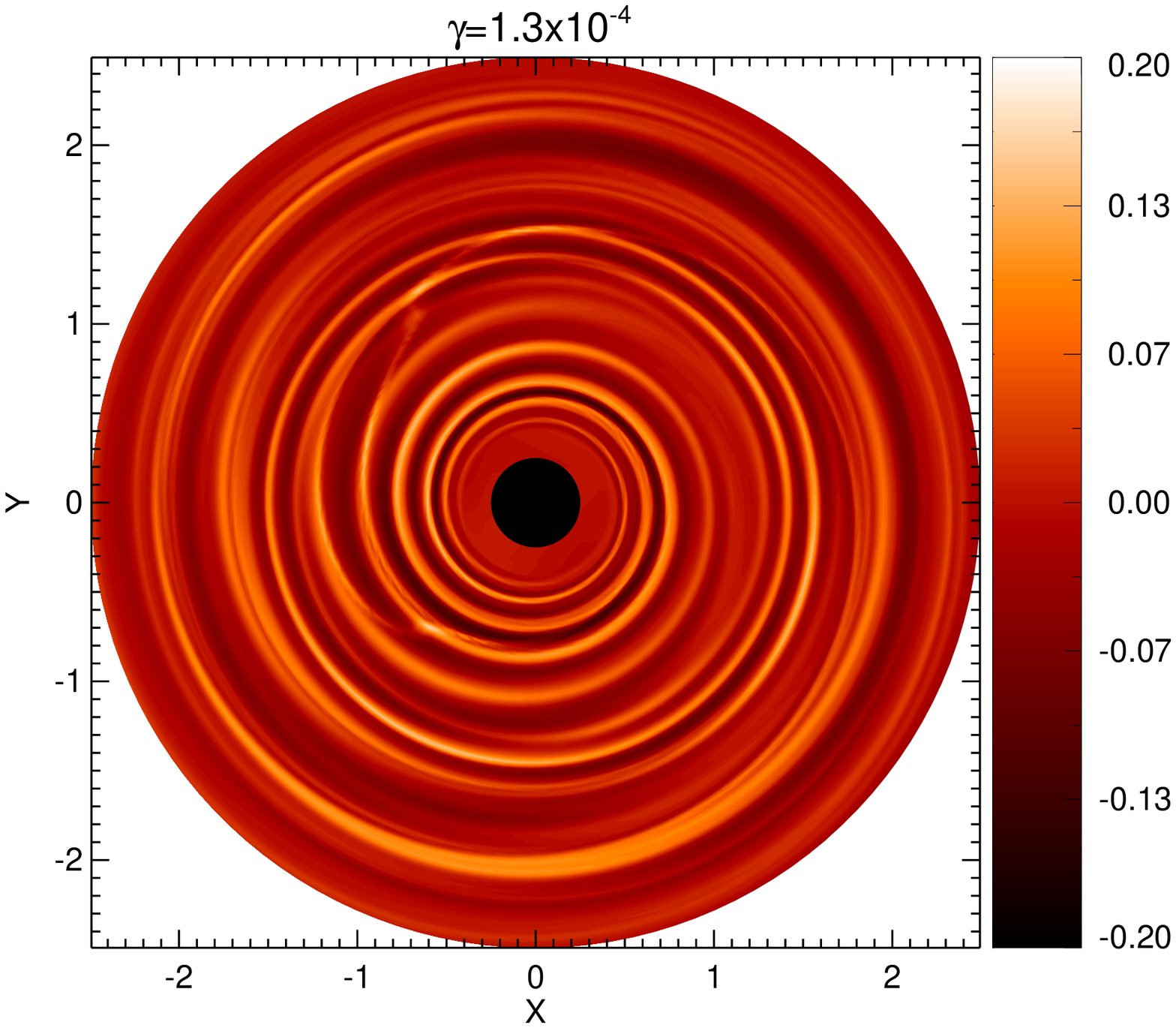}
\includegraphics[width=0.8\columnwidth]{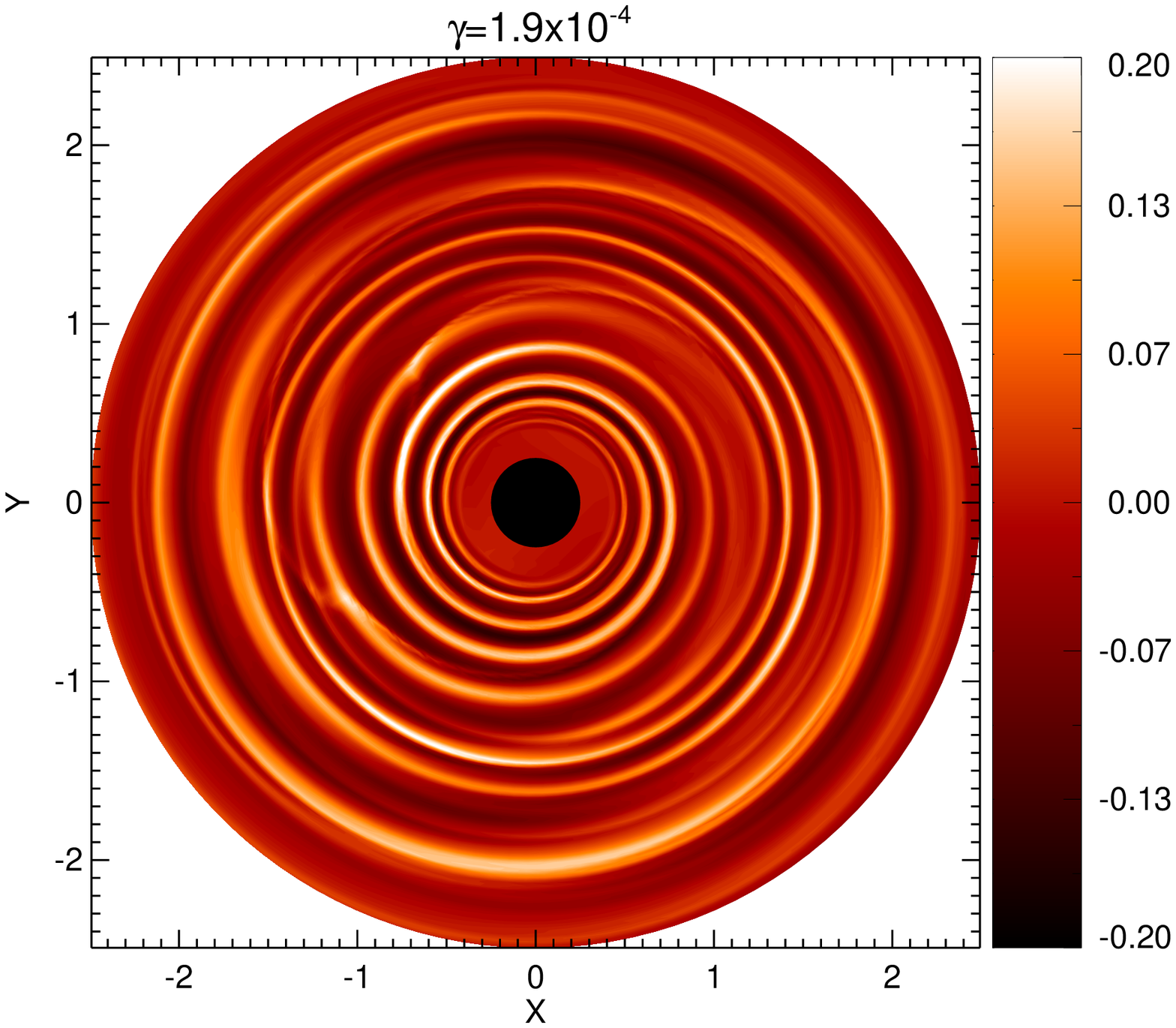}
\caption{This figure shows, for model G1, snapshots of the perturbed surface density of the disk for $\gamma=0$ (first panel), 
$\gamma=6\times 10^{-5}$ (second panel), $\gamma=1.3\times 10^{-4}$ (third panel) and 
$\gamma=1.9\times 10^{-4}$ (fourth panel).}
\label{sig2d}
\end{figure*}
 
\begin{figure}
\centering
\includegraphics[width=0.9\columnwidth]{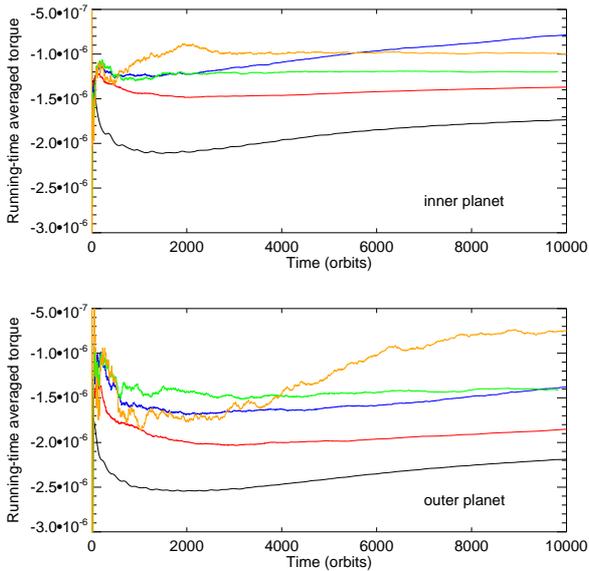}
\caption{{\it Upper panel:} time evolution of the running-time 
averaged torques 
exerted on the inner planet for model G1 and for the different values
of $\gamma$ we considered namely for
$\gamma=0$ (black line), $\gamma=6\times 10^{-5}$ (red line), $\gamma=1.3\times 10^{-4}$ (blue),
$\gamma=1.9\times 10^{-4}$ (green) and $\gamma=3\times 10^{-4}$ (orange). 
{\it Lower panel:} same but for the outer planet.}
\label{torque}
\end{figure}
\begin{figure}
\centering
\includegraphics[width=0.9\columnwidth]{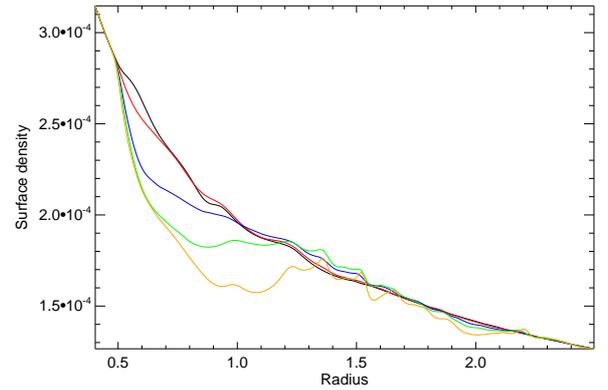}
\caption{Disk surface density profile at $t=2000$ orbits for model G1 and for the different values
of $\gamma$ we considered namely for
$\gamma=0$ (black line), $\gamma=6\times 10^{-5}$ (red line), $\gamma=1.3\times 10^{-4}$ (blue),
$\gamma=1.9\times 10^{-4}$ (green) and $\gamma=3\times 10^{-4}$ (orange).}
\label{sig2000}
\end{figure}
The time evolution of the planets' semi-major axes, eccentricities and period ratio corresponding 
to model G1 and for one realization of the different values of $\gamma$ we considered 
are depicted in Fig. \ref{run1}. In each case, the period ratio is observed to initially decrease, suggesting
that early evolution involves convergent migration of the two planets. Not surprisingly,
a tendancy for the planets to undergo a monotonic inward migration is observed at the beginning of the 
simulations with the lowest values of $\gamma$ whereas these are more influenced
by stochastic forcing for $\gamma \ge 1.9 \times 10^{-4}$. This is due to the fact that the amplitude of 
the turbulent density fluctuations is typically stronger than that of the planet's wake for simulations
with $\gamma \ge 1.9\times 10^{-4}$, as shown in Fig. \ref{sig2d} which displays snapshots of the 
 perturbed surface density of the disk for different values of $\gamma$.\\
 The semimajor axis 
evolution also reveals a clear tendancy for lower 
migration rates with increasing $\gamma$, which is an effect arising from 
the desaturation of the horseshoe drag by turbulence (Baruteau \& Lin 2010). As $\gamma$ increases, 
 the disk torques are indeed expected to increase from the differential Lindblad torque, obtained for $\gamma=0$, 
up to the fully unsaturated torque, which is maintained for $\alpha\sim 0.16 (m_i/M_*)^{3/2}h^{-4}$ 
(Baruteau \& Lin 2010). For $h=0.05$, such a value for $\alpha$ corresponds to 
$\gamma \sim 1.2\times 10^{-4}$. For higher values of $\gamma$ however, we expect the torques 
to slightly decrease with increasing $\gamma$ due to a cut-off of the horseshoe drag arising
when the diffusion timescale across the horseshoe region is smaller than the horseshoe U-turn 
time (Baruteau \& Lin 2010). \\
 This can be confirmed by inspecting the  evolution of the running-time averaged 
torques exerted on both planets and which 
are presented in Fig. \ref{torque}. Up to a time of approximately $\sim 10^3$ orbits and for 
$\gamma \le 1.3\times 10^{-4}$, the torques are 
observed to increase with increasing $\gamma$,  while they decrease for higher values of $\gamma$. This is 
in  very good agreement with the expectation that the fully unsaturated torque is reached for 
$\gamma=1.2\times 10^{-4}$. At later times however, the torques obtained in simulations with $\gamma \ge 1.9 \times 10^{-4}$ 
 can eventually exceed those computed in runs with $\gamma \le 1.3\times 10^{-4}$. We suggest that this is
 related to the fact 
that the disk surface density tends to be significantly modified at the planet positions at high turbulence level. 
This is illustrated in Fig. \ref{sig2000} which
shows the disk surface density profile at $t=2000$ orbits for the different values of $\gamma$ we considered. Here the inner and 
outer planets are located at $a_i\sim 0.98$ and $a_o\sim 1.25$ respectively. It is interesting to note that for
 $\gamma=3\times 10^{-4}$, the outer planet tends to
evolve in a region of positive surface density gradient where the corotation torque is positive, in such a way 
that the torque exerted on that planet can become higher than that exerted on the inner one. This effect is 
responsible  for the increase of period ratio observed at late times in simulations with 
$\gamma \ge 1.9 \times 10^{-4}$.

\subsubsection{Eccentricity evolution}
\label{sec:eccentricity}

\begin{figure}
\centering
\includegraphics[width=0.9\columnwidth]{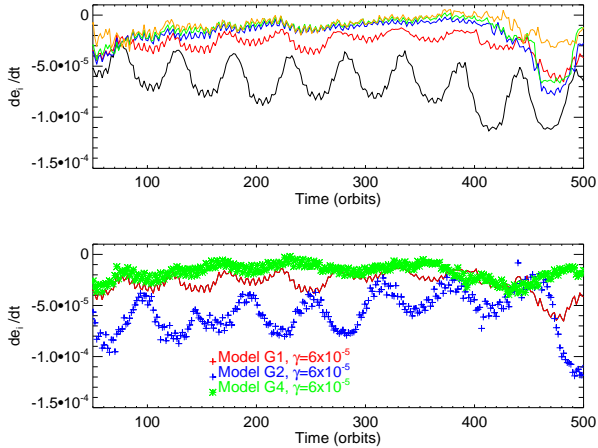}
\caption{{\it Upper panel:} time evolution of the theoretical 
change of the inner planet eccentricity $de_i/dt$ given by Eq. \ref{eq:dei} for model G1 and for the different values
of $\gamma$ we considered namely for
$\gamma=0$ (black line), $\gamma=6\times 10^{-5}$ (red line), $\gamma=1.3\times 10^{-4}$ (blue),
$\gamma=1.9\times 10^{-4}$ (green) and $\gamma=3\times 10^{-4}$ (orange). {\it Lower panel:} 
same but for $\gamma=6\times 10^{-5}$ and models G1 (red), G2 (blue) and G4 (green).} 
\label{eccplot}
\end{figure}

 For model G1, examination of the early planets' eccentricities evolution displayed in Fig. \ref{run1} 
shows a clear tendancy of higher eccentricities with 
increasing the value for $\gamma$, which is in agreement with the expectation that turbulence is a source of eccentricity driving
(Nelson 2005). This can be confirmed by inspecting the theoretical change of the inner planet eccentricity $de_i/dt$ 
which can be computed using the following expression (Burns 1976):

\begin{equation}
\frac{de_i}{dt}=\frac{e_i^2-1}{2e_i^2}\left(\frac{\dot E}{E}+2\frac{\dot H}{H}\right),
\label{eq:dei}
\end{equation} 

where $E=-G(M_\star+m_i)/2a_i$ is the specific energy of the inner planet, $H=\sqrt{G(M_\star+m_i)a_i(1-e_i^2)}$
its specific angular momentum, $\dot H$ the torque exerted by the disk and $\dot E$ the power of the
force exerted by the disk on the planet. The early time evolution of $de_i/dt$ is displayed, for this model and
for the different values of $\gamma$ we considered  in the upper panel of Fig. \ref{eccplot}.  It clearly demonstrates 
that, compared with the laminar run,  the theoretical rate of change of $e_i$ is higher in turbulent runs and that it
increases with increasing the value for $\gamma$. \\
Also, inspection of the lower panel  of Fig. \ref{eccplot} reveals that, compared with model G1 
in which $m_i=m_o=3.3 \; M_\oplus$ and $\Sigma_0=2\times  10^{-4}$, the disk induced 
eccentricity damping is weaker in model G4 for which 
$m_i=m_o=1.6\; M_\oplus$.  This is due to the
damping of eccentricity at coorbital Lindblad resonances scaling linearly with planet mass (Nelson 2005). Given that this damping
also scales with disk mass, this explains why the disk induced eccentricity damping is stronger in model G2 in which 
$\Sigma_0=4\times  10^{-4}$.

\subsubsection{Time evolution of the resonant angles}
\begin{figure}
\centering
\includegraphics[width=0.9\columnwidth]{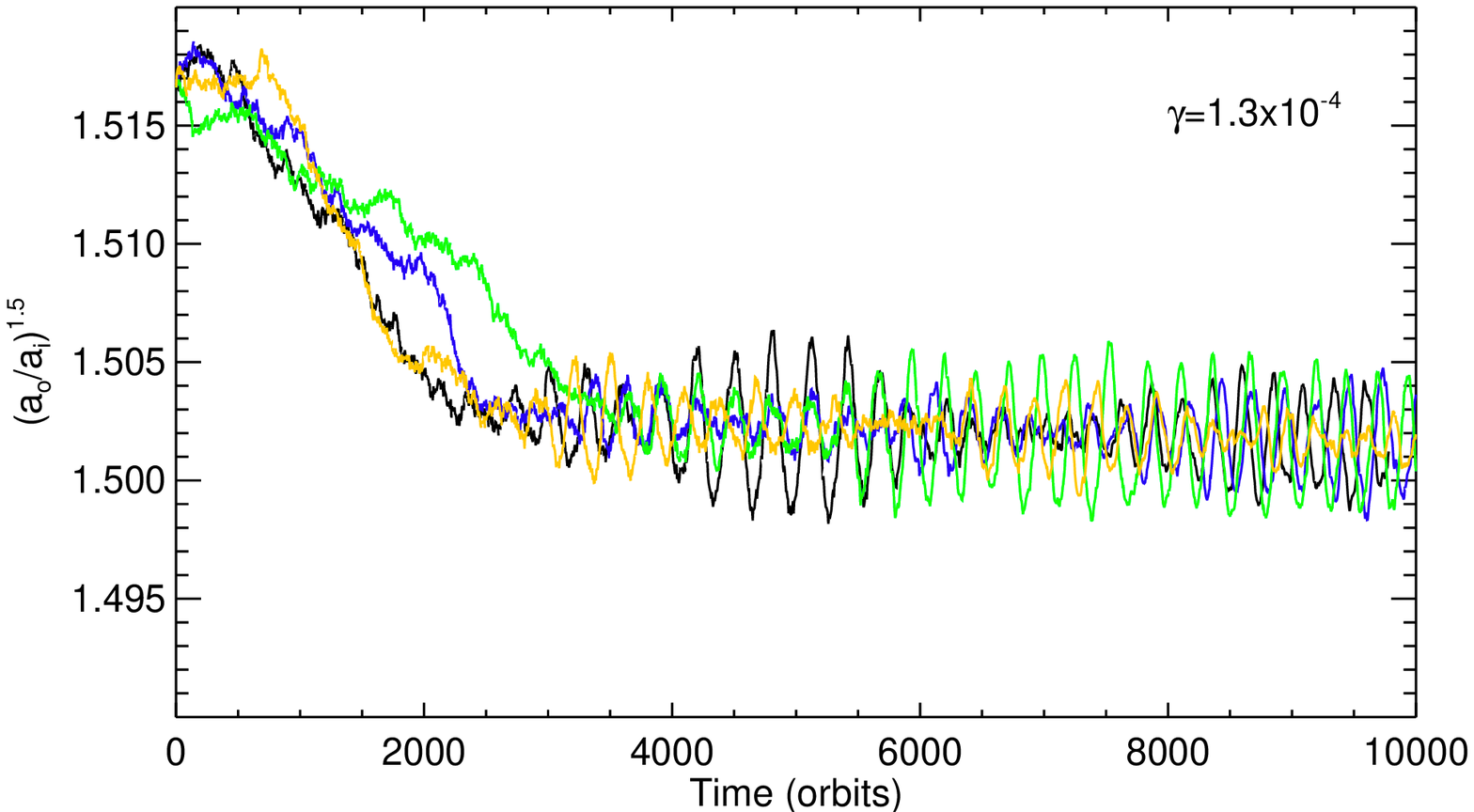}
\includegraphics[width=0.9\columnwidth]{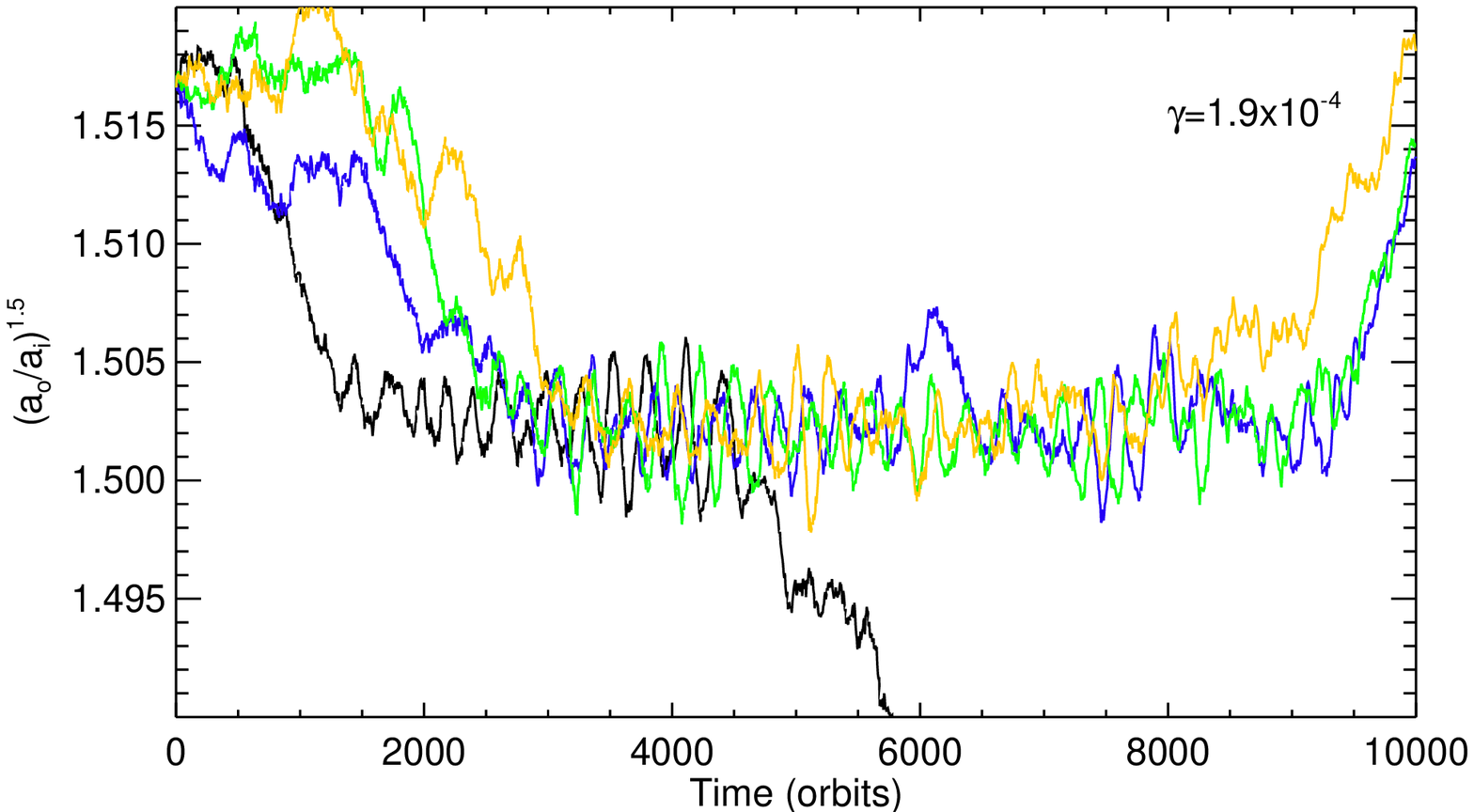}
\caption{{\it Upper panel:} time evolution of the period ratio for model G1 and for four 
different realizations with $\gamma=1.3\times 10^{-4}$. {\it Lower panel:} same but for 
$\gamma=1.9\times 10^{-4}$. Simulations were performed with GENESIS.}
\label{stat}
\end{figure}
\begin{figure}
\centering
\includegraphics[width=\columnwidth]{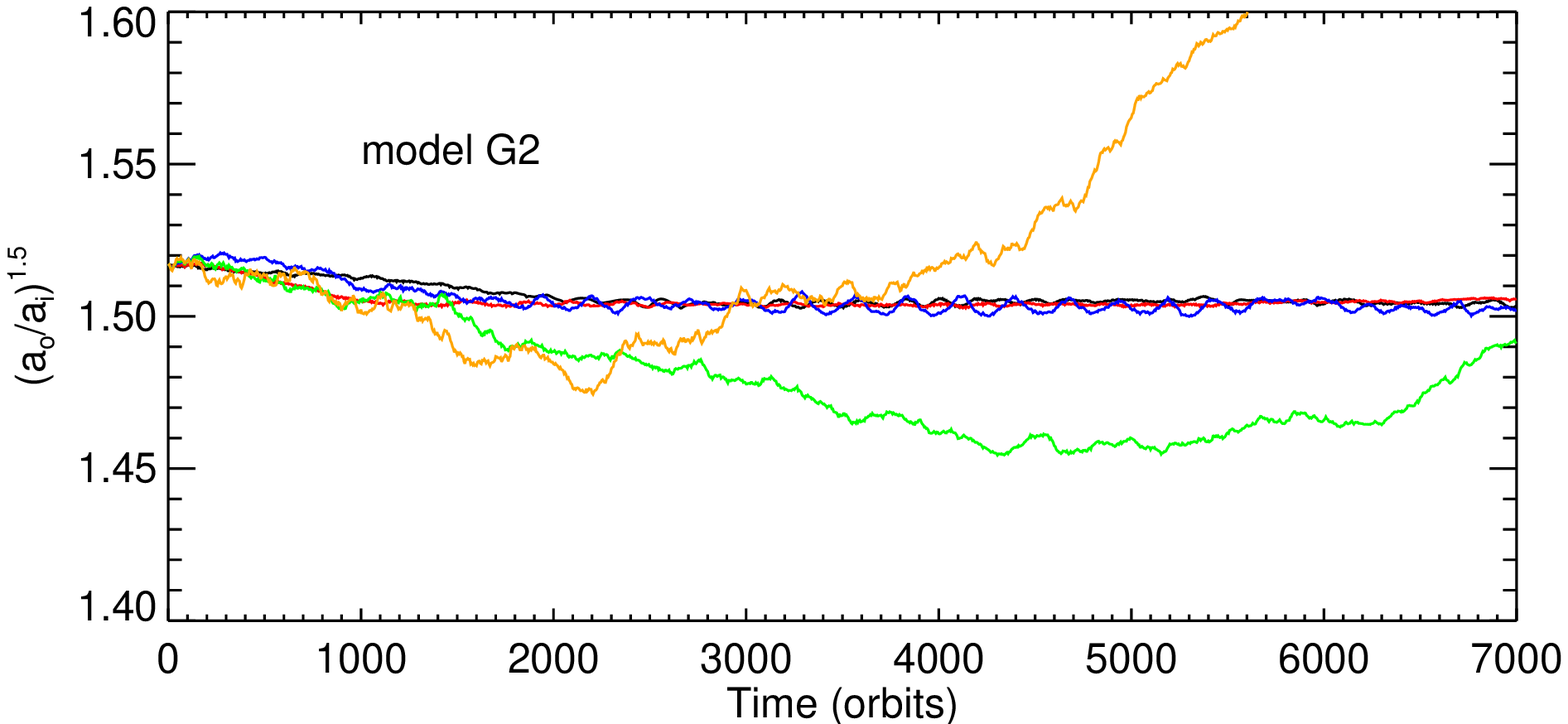}
\includegraphics[width=\columnwidth]{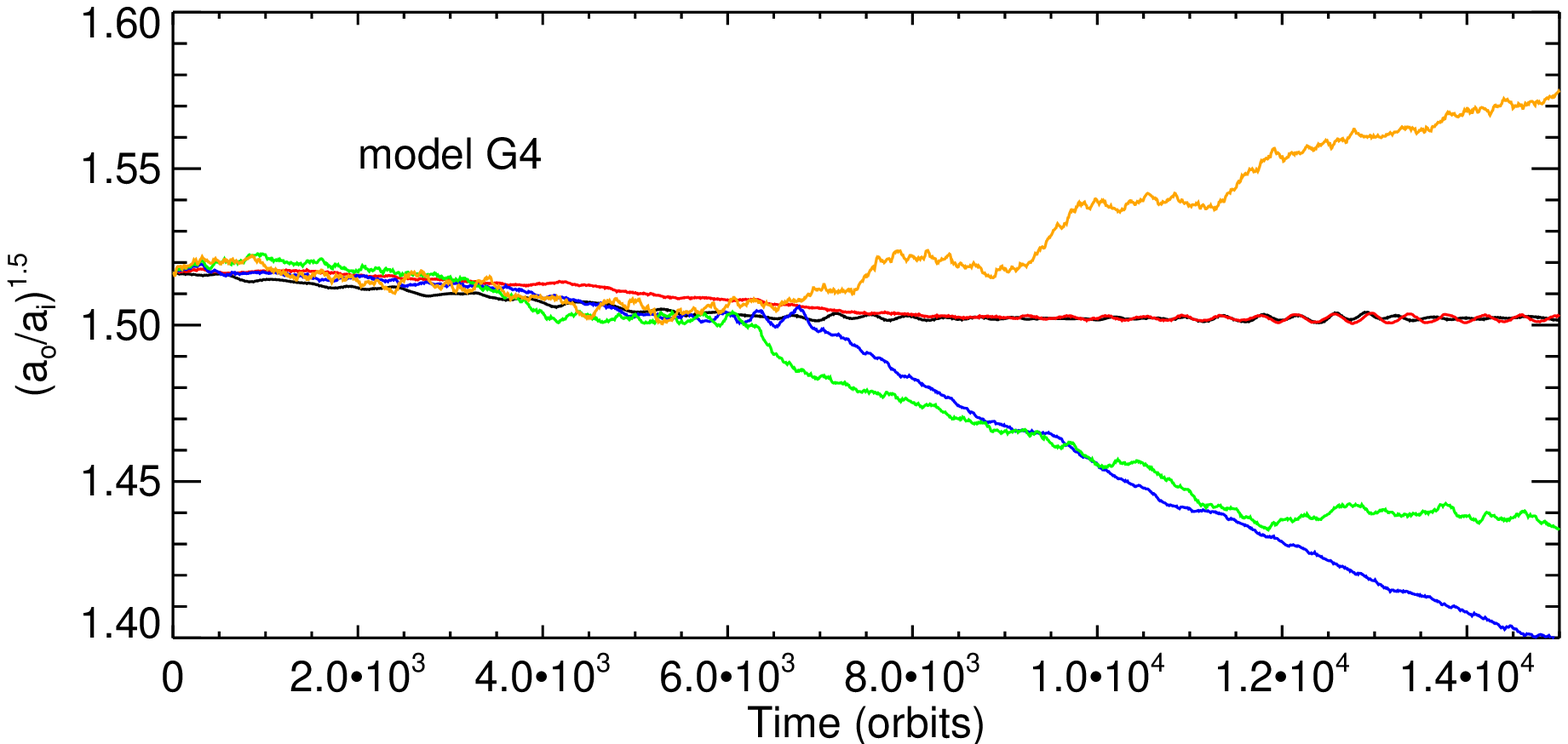}
\caption{{\it Upper panel:} time evolution of the period ratio  for 
model G2 and for the different values
of $\gamma$ we considered namely for
$\gamma=0$ (black line), $\gamma=6\times 10^{-5}$ (red), $\gamma=1.3\times 10^{-4}$ (blue),
$\gamma=1.9\times 10^{-4}$ (green) and $\gamma=3\times 10^{-4}$ (orange). {\it Lower panel:} same but for 
model G4. Simulations were performed with GENESIS.}
\label{per}
\end{figure}

\begin{figure}
\centering
\includegraphics[width=\columnwidth]{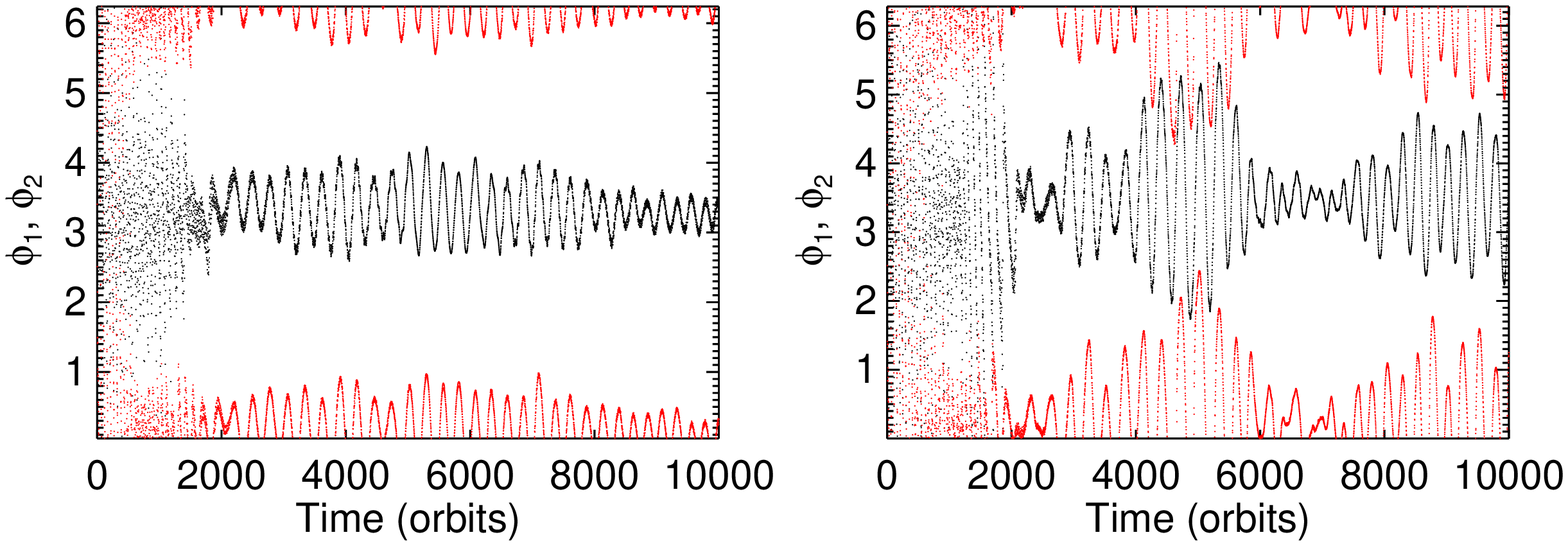}
\includegraphics[width=\columnwidth]{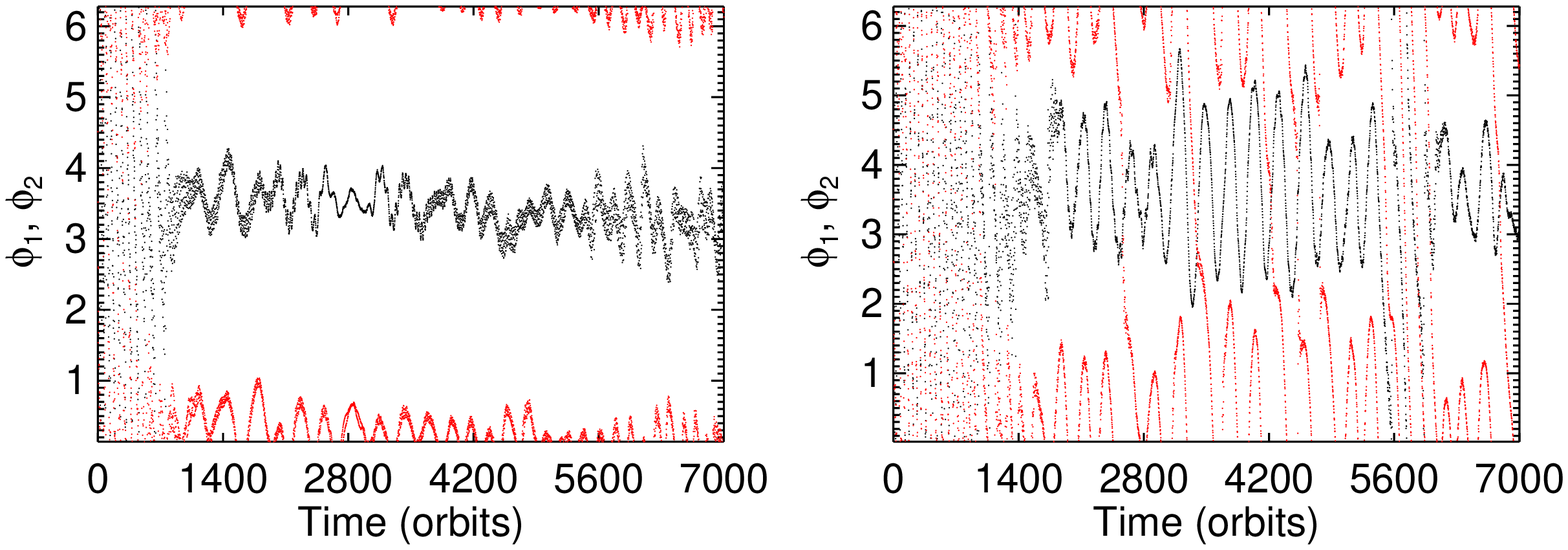}
\includegraphics[width=\columnwidth]{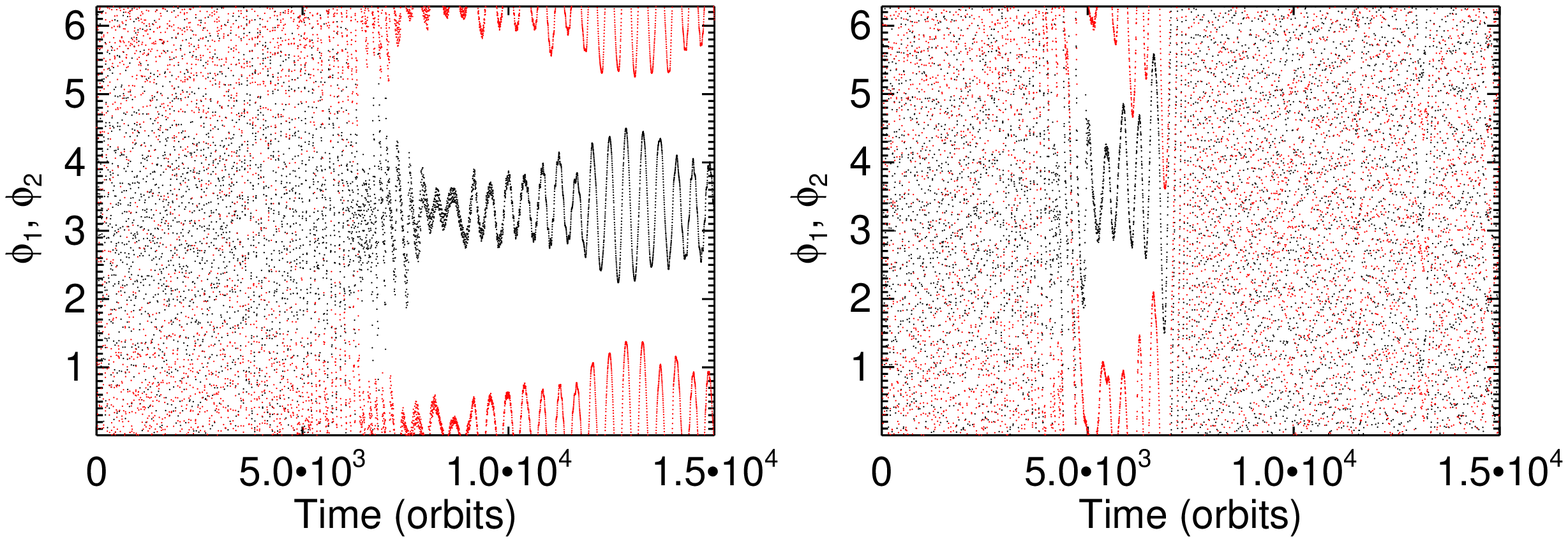}
\caption{{\it Upper panel:} time evolution of the 
resonant angles $\phi_1=3\lambda_o-2\lambda_i-\omega_i$ (black) and 
$\phi_2=3\lambda_o-2\lambda_i-\omega_o$ (red) for model G1 with $\gamma=6\times 10^{-5}$ (left) and 
$\gamma=1.3\times 10^{-4}$ (right). {\it Middle panel:} same but for model $G2$. 
{\it Lower panel:} same but for model G4.}  
\label{angles}
\end{figure}
 As mentioned above, we find that for model G1 a stable 3:2 
commensurability generally forms for $\gamma \le 1.3 \times 10^{-4}$, which can be confirmed by inspecting the 
upper panel of Fig. \ref{stat} which displays the time 
evolution of the period ratio for four different realizations with this value of $\gamma$. However, as one can see in the 
lower panel of Fig. \ref{stat}, such a resonance  
is observed to break at late times for all the realizations performed with $\gamma \ge 1.9\times 10^{-4}$.  Comparing 
Fig. \ref{stat} and the two upper panels in Fig. \ref{nbody}, we see that the results of these hydrodynamical 
simulations are at first sight in good agreement with those of N-body runs. Moreover, it is interesting to note that 
in some runs, planets which leave the 3:2 resonance can eventually pass through 
that resonance again at later times 
or others commensurabilities like the 2:1 resonance. \\
A similar outcome is observed in model G2 which had
a value of $\Sigma_0=4\times 10^{-4}$.  This arises because both the turbulent torque and the damping rate of the 
libration amplitude scale linearly with $\Sigma_0$. For model G4 however, which had $\Sigma_0=2\times 10^{-4}$ and $m_i=m_o=1.6\;M_\oplus$, the 3:2 resonance is
disrupted in the run with $\gamma=1.3\times 10^{-4}$ due to the damping rate being smaller for this model. For a single 
realization for each value of $\gamma$, the evolution of the period ratio for  models G2 and G4 is displayed in 
Fig. \ref{per}. \\


In the two upper panels of Fig. \ref{angles} we show, for 
model G1 and for  two realizations with $\gamma=6\times 10^{-5}$ and $\gamma=1.3\times 10^{-4}$, the time 
evolution of the resonant angles associated with the 3:2 resonance. 
For $\gamma=6\times 10^{-5}$, 
 the 3:2 resonance is established  at $t\sim 1800$ orbits while it forms at $t\sim 2000$ orbits
for the calculation with $\gamma=1.3\times 10^{-4}$. This is consistent with the fact that, for moderate values 
of $\gamma$, migration rates tend to decrease with increasing $\gamma$. As illustrated in the second 
and third panels of Fig. \ref{run1}, 
resonant capture 
makes the eccentricities of both
planets grow rapidly  before they saturate at values of $e_i\sim e_o\sim 0.01$ in the run with 
$\gamma=6\times 10^{-5}$. As discussed in  Sect. \ref{sec:eccentricity},  these tend to be larger in the case 
where $\gamma=1.3\times 10^{-4}$, with the eccentricities
reaching peak values of $e_i\sim 0.02$ and $e_o\sim 0.015$. \\
Not surprisingly, there is a clear trend for the  amplitude of the resonant angles 
 to increase with increasing the value for $\gamma$  in model G1. For the run  with $\gamma=6\times 10^{-5}$, 
the angles slightly spread until $t\sim 5\times 10^3$ orbits  before their amplitude continuously decrease with
time. This  
 indicates that over long timescales damping of the resonant angles through migration tends to overcome diffusion effects.  In the  
case where $\gamma =1.3\times 10^{-4}$ however,  periods of cyclic variations of the resonant angles can be seen  
with the angles librating with high amplitude before being subsequently damped. Given that in absence of turbulent 
forcing, the libration amplitude should decrease as $\Omega_i^{-1/2}$ (Peale 1976), we would expect 
the 3:2 resonance to be maintained, for  $\gamma \le 1.3\times 10^{-4}$, over timescales much longer than 
those covered by the simulations.\\
 For two realizations with $\gamma=6\times 10^{-5}$ and $\gamma= 1.3\times 10^{-4}$, the evolution of both $\phi_1$ and $\phi_2$ for  
models  G2 and G4 is diplayed 
in the middle and lower panels of Fig. \ref{angles} respectively. 
In comparison with model G1, the resonant angles librate with slightly higher amplitudes in model G2 and can 
eventually start 
oscillations between periods of 
circulation and libration in the run with $\gamma=1.3\times 10^{-4}$.  Again this arises because, compared with model 
G1, the turbulent density
fluctuations  are stronger in this model. In model G4 however,  the 3:2 
resonance is maintained for only $\sim 3\times 10^3$ orbits in the case where $\gamma=1.3\times 10^{-4}$, which indicates that for this model the damping rate is 
too weak for this resonance to remain stable.     
\begin{figure}
\centering
\includegraphics[width=0.9\columnwidth]{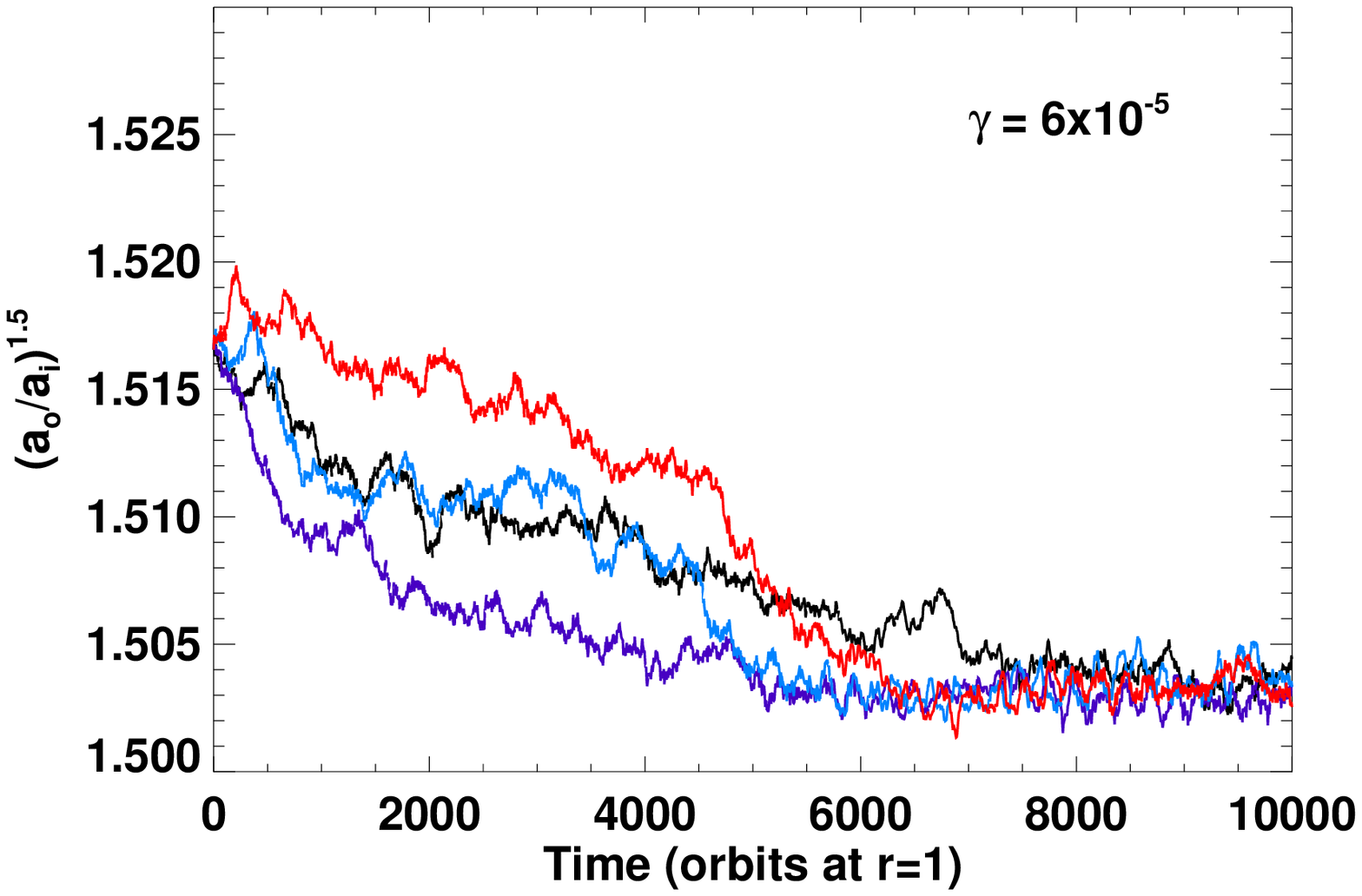}
\includegraphics[width=0.9\columnwidth]{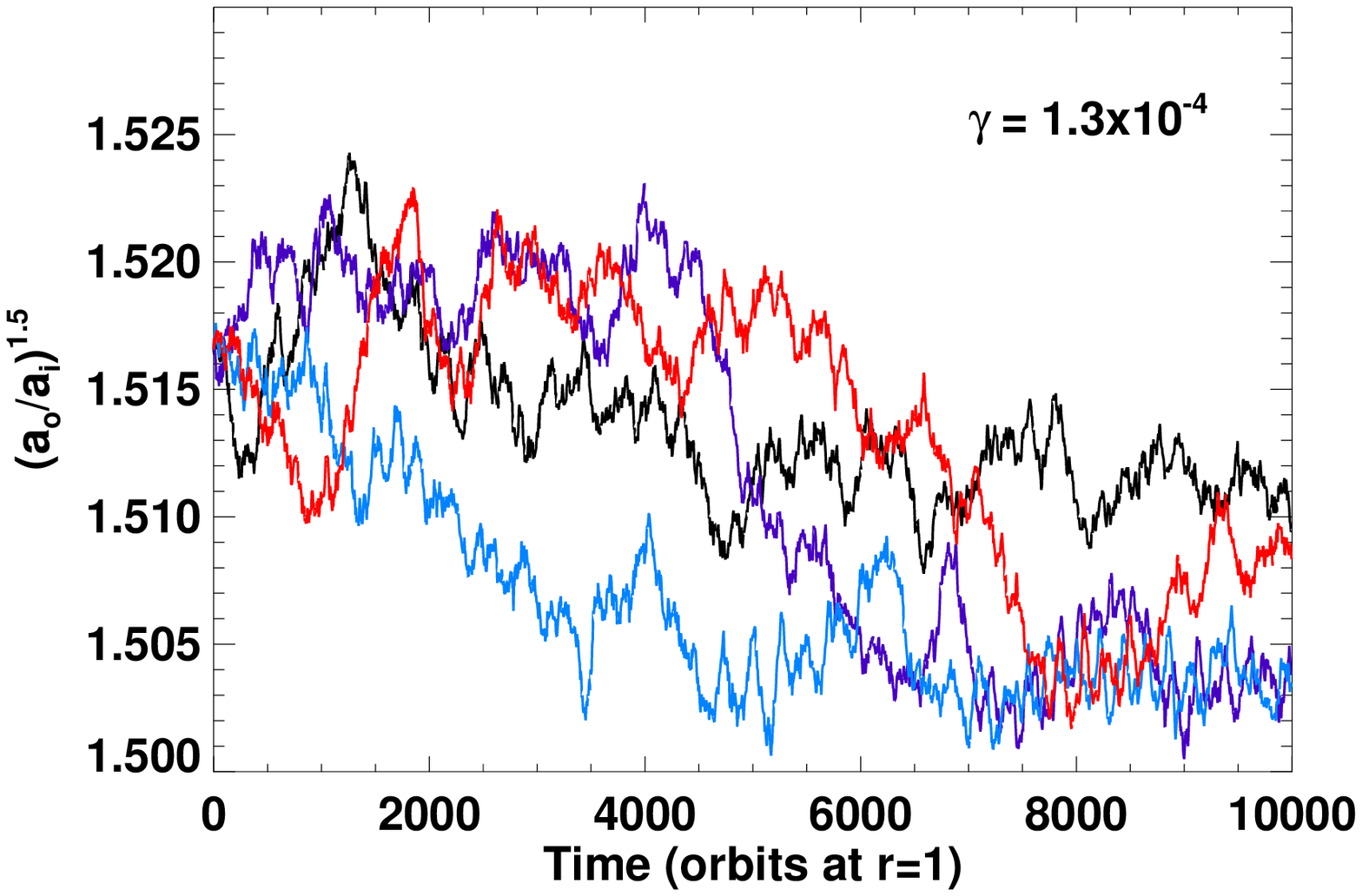}
\caption{{\it Upper panel:} time evolution of the period ratio for model G1 and for four 
runs with $\gamma=6\times 10^{-5}$ and in which Eq. \ref{eq:damp} is solved at each timestep . {\it Lower panel:} same but for 
$\gamma=1.3\times 10^{-4}$. Simulations were performed with FARGO.}
\label{damp}
\end{figure}

\begin{figure}
\centering
\includegraphics[width=0.49\columnwidth]{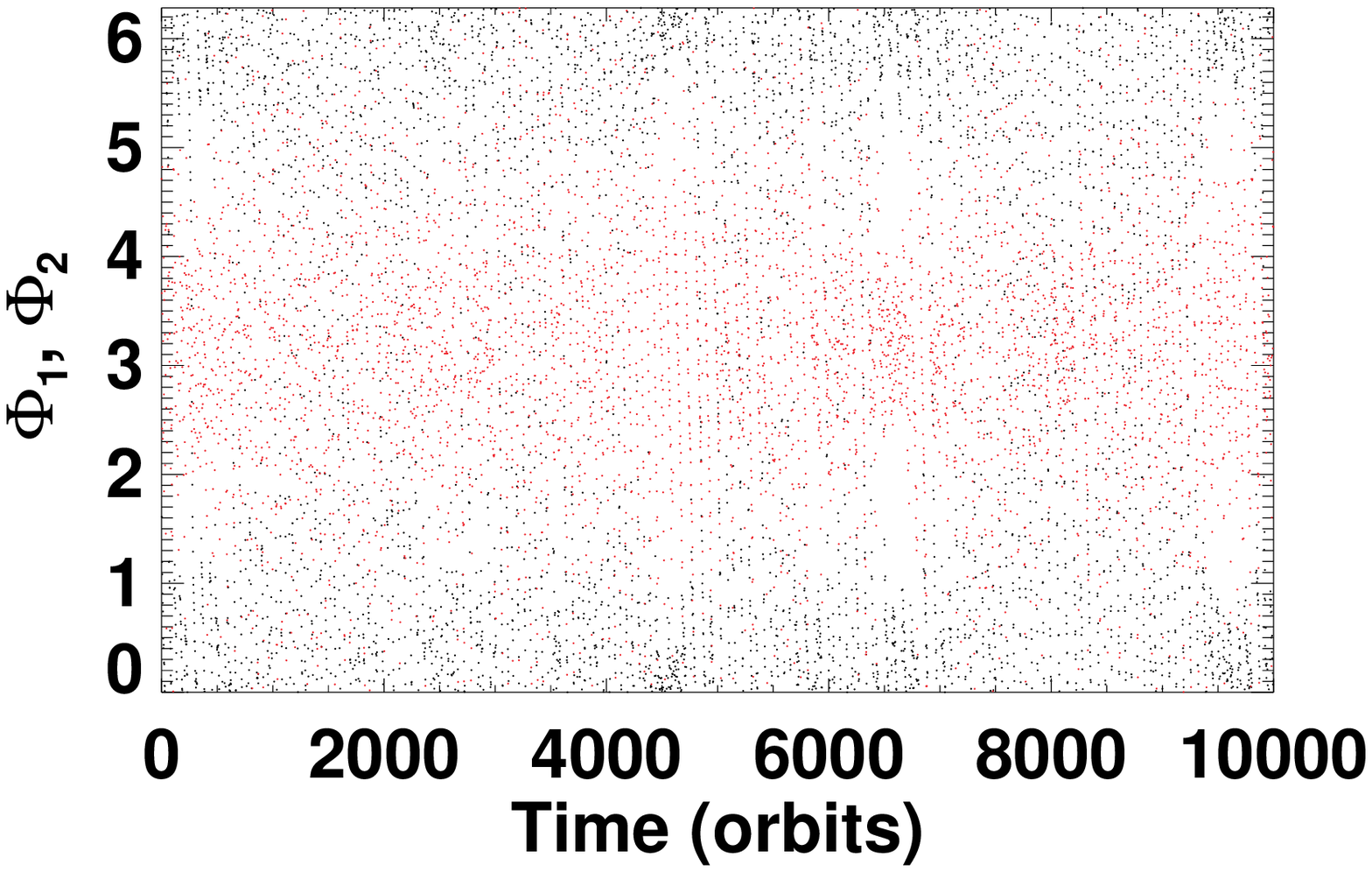}
\includegraphics[width=0.49\columnwidth]{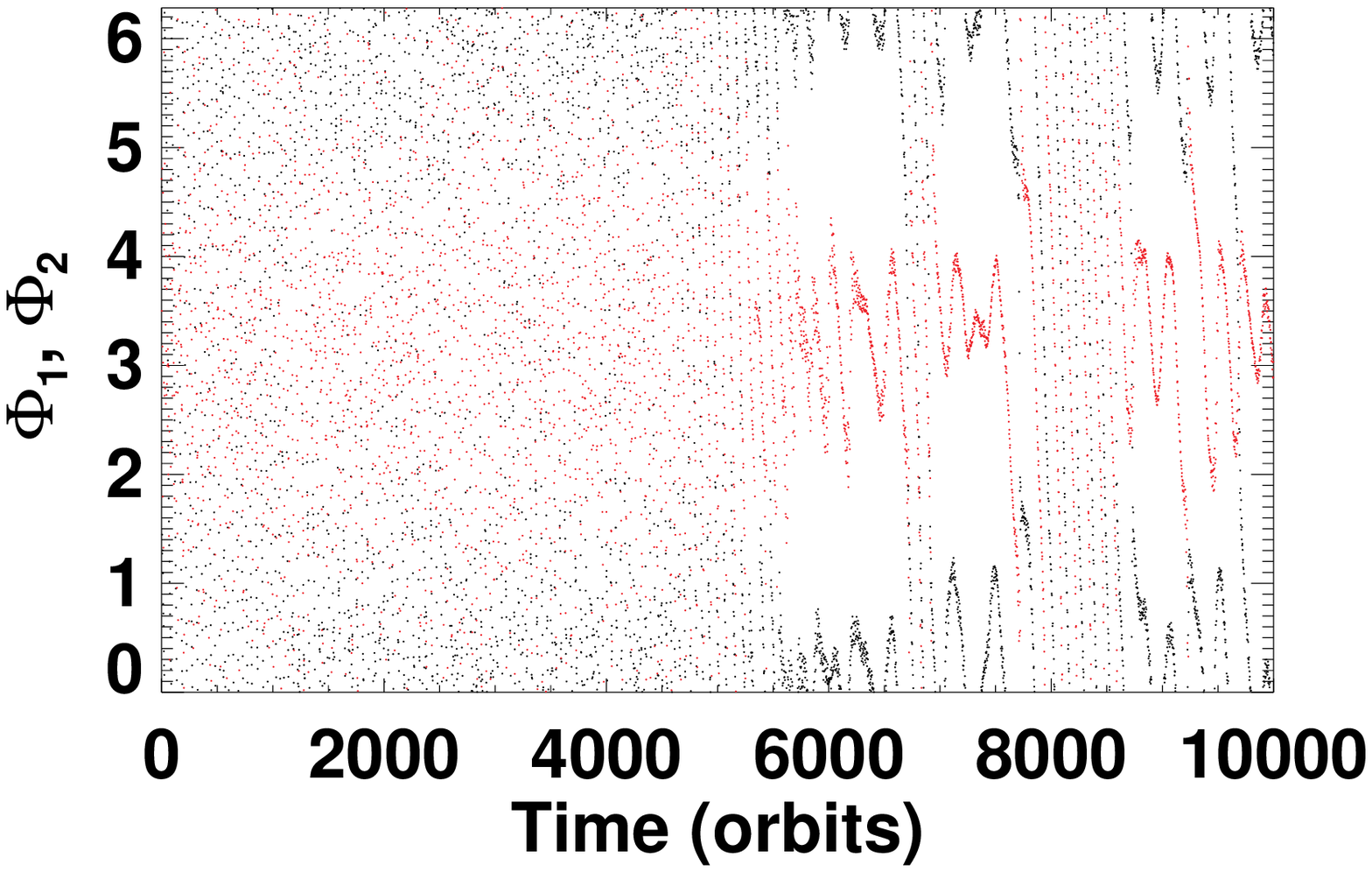}
\includegraphics[width=0.49\columnwidth]{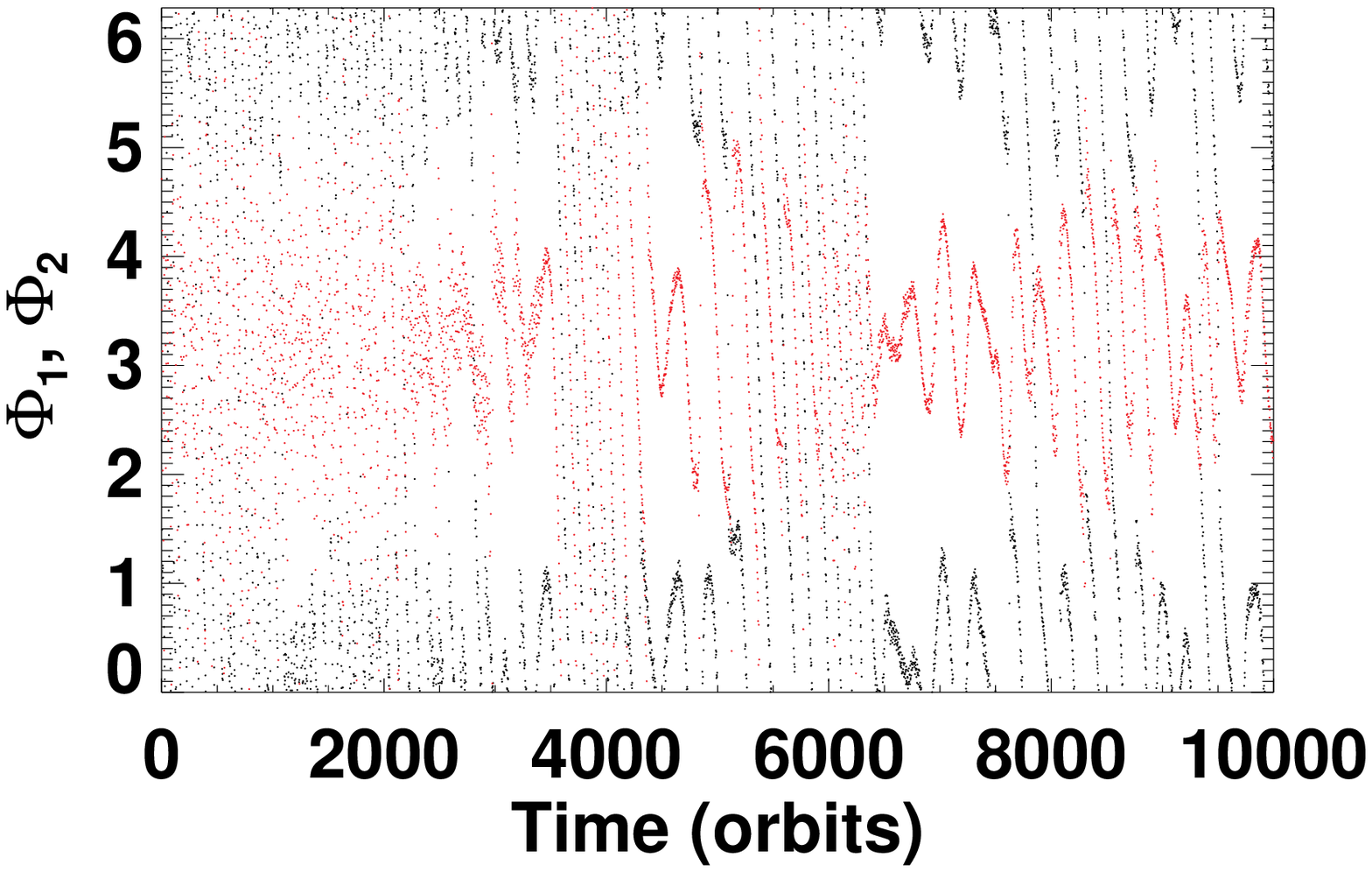}
\includegraphics[width=0.49\columnwidth]{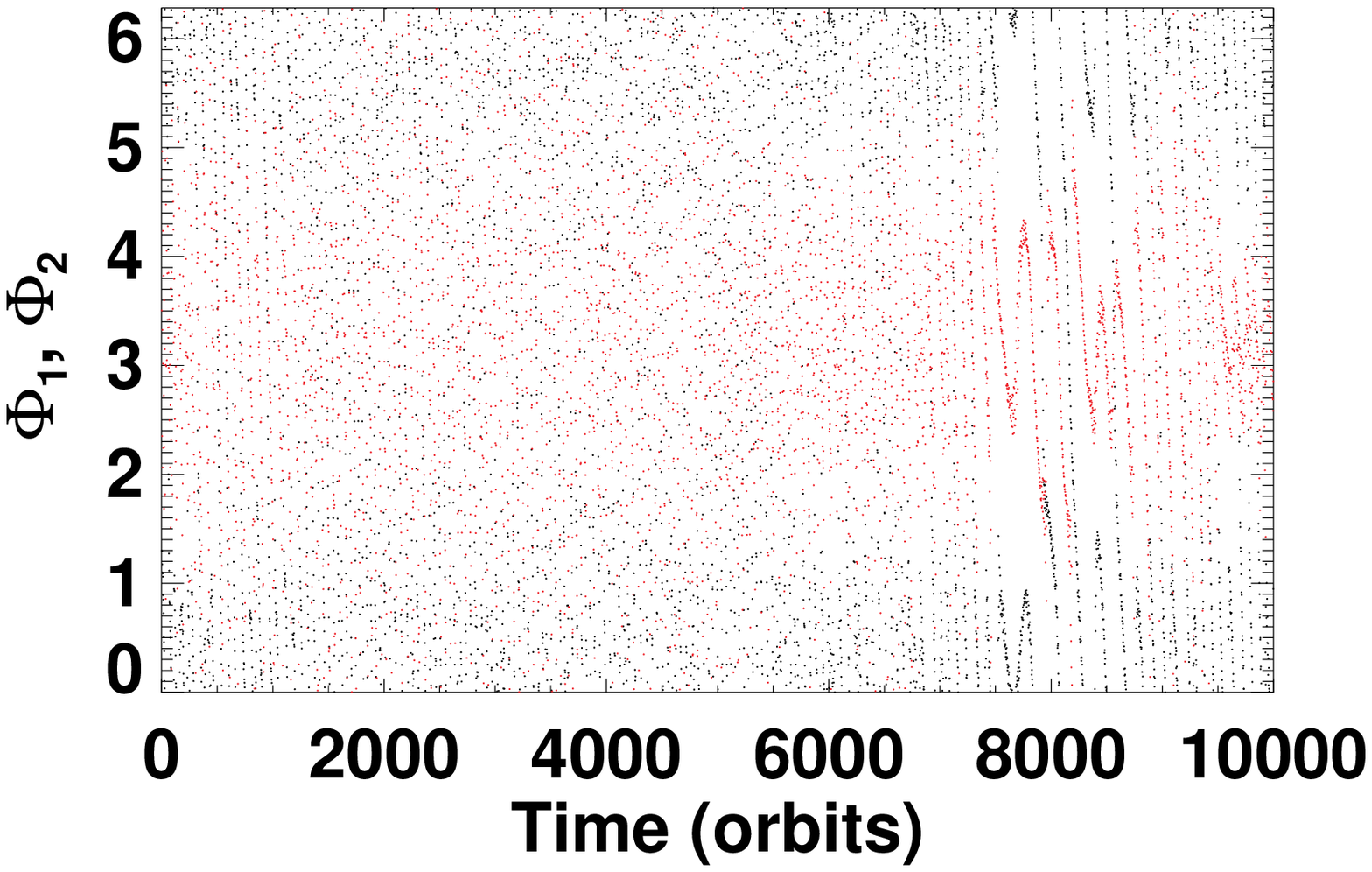}
\caption{Time evolution, for model G1 and for four different 
realizations, of the 
resonant angles $\phi_1=3\lambda_o-2\lambda_i-\omega_i$ (black) and 
$\phi_2=3\lambda_o-2\lambda_i-\omega_o$ (red) for  
$\gamma=1.3\times 10^{-4}$ and in the case where Eq. \ref{eq:damp} is solved at each timestep.}  
\label{angleswithdamping}
\end{figure}
\subsubsection{Comparison with analytics}
\label{sec:comparison}

 We now examine how the results of the simulations described above compare with the expectations discussed 
in Sect. \ref{sec:exp}. For model G1, we can estimate the 
 libration frequency $\omega_0 $ using 
Eq. \ref {eq:frequency} in conjunction with the results from this model shown in Fig. \ref{run1}.  Adopting the 
inviscid simulation as a fiducial case, we have $a_i=0.9$, $a_o=1.17$ and $e_i=0.01$ at $t\sim 4000$ orbits, which leads 
to $\delta \omega=\omega_0/\Omega_o \sim 1.9\times 10^{-3}$. Moreover, the migration
timescale {of the system} was estimated to $\tau_{mig}\sim 3.3\times 10^4$  orbits from the results of this simulation.  
  Using Eq. \ref{eq:gammac} which gives the value for $\gamma_c$ as a function of $\tau_{mig}$, 
we can then provide an analytical estimate $\gamma_c^{ana}$ of the critical 
amplitude for the turbulent forcing above which the 3:2 resonance should be disrupted. For this model, this critical value 
is estimated to be  $\gamma_c^{ana}\sim 1.9\times 10^{-4}$  while Eq. \ref{eq:gammac2} predicts 
$\gamma_c^{ana}\sim 2.5\times 10^{-4}$. Returning to Fig. \ref{stat}, we see that the results of the 
simulations performed with GENESIS suggest 
that $1.3\times 10^{-4}\le\gamma_c<1.9\times 10^{-4}$ for this model, which is clearly in broad agreement with the 
previous analytical estimate.  We note however that both simulations performed with FARGO and additional runs in which a 
roughly constant surface density 
profile is maintained (see Sect. \ref{sec:num}) produced slightly different results since 
we find $6\times 10^{-5}\le\gamma_c<1.3\times 10^{-4}$ in these cases. The little difference exhibited by our two codes is apparently 
due to the fact that turbulence induces changes in the surface density profile which are slightly different. In FARGO, 
the disk density at the position of the inner planet is slightly higher compared with GENESIS while it is slightly lower at the 
position of the outer planet.\\
 For calculations in which Eq. \ref{eq:damp} is 
solved at each timestep, the time evolution of the period ratio for 
four realizations with $\gamma=6\times 10^{-5}$ and $\gamma=1.3\times 10^{-4}$ is displayed in Fig. \ref{damp}.
  In that case, all the realizations performed with $\gamma=6\times 10^{-5}$ resulted in the formation of the 3:2 resonance whereas
for $\gamma=1.3\times 10^{-4}$, two of the four realizations resulted in capture in that resonance by the end of the run. For 
$\gamma=1.3\times 10^{-4}$, the time evolution of the resonant angles associated with the 
3:2 resonance  is displayed in Fig. \ref{angleswithdamping}. Compared with previous runs in which the surface density profile 
was altered by turbulence, we see that capture in resonance tends to occur later in runs where a roughly constant surface density profile is maintained. This occurs because the disk density at the position of the inner planet tends to be higher in runs where the initial 
disk surface density 
profile is restored, leading to a slower differential migration between the two planets.    \\
For other models, repeating the previously decribed procedure leads to analytical estimates of $\gamma_c^{ana}=1.2\times 10^{-4}$ for model 
G2 and $\gamma_c^{ana} \sim 9.3\times 10^{-5}$ for model G4. Given that the simulations performed with GENESIS 
indicate that $1.3\times 10^{-4}\le\gamma_c<1.9\times 10^{-4}$ and $6\times 10^{-5}\le\gamma_c<1.3\times 10^{-4}$ 
for models G2 and G4 respectively, we see that again the 
previous analytical estimates compare reasonably well with the results of our simulations.

\subsection{Model with $q=1/2$}
\begin{figure*}
\centering
\includegraphics[width=0.8\textwidth]{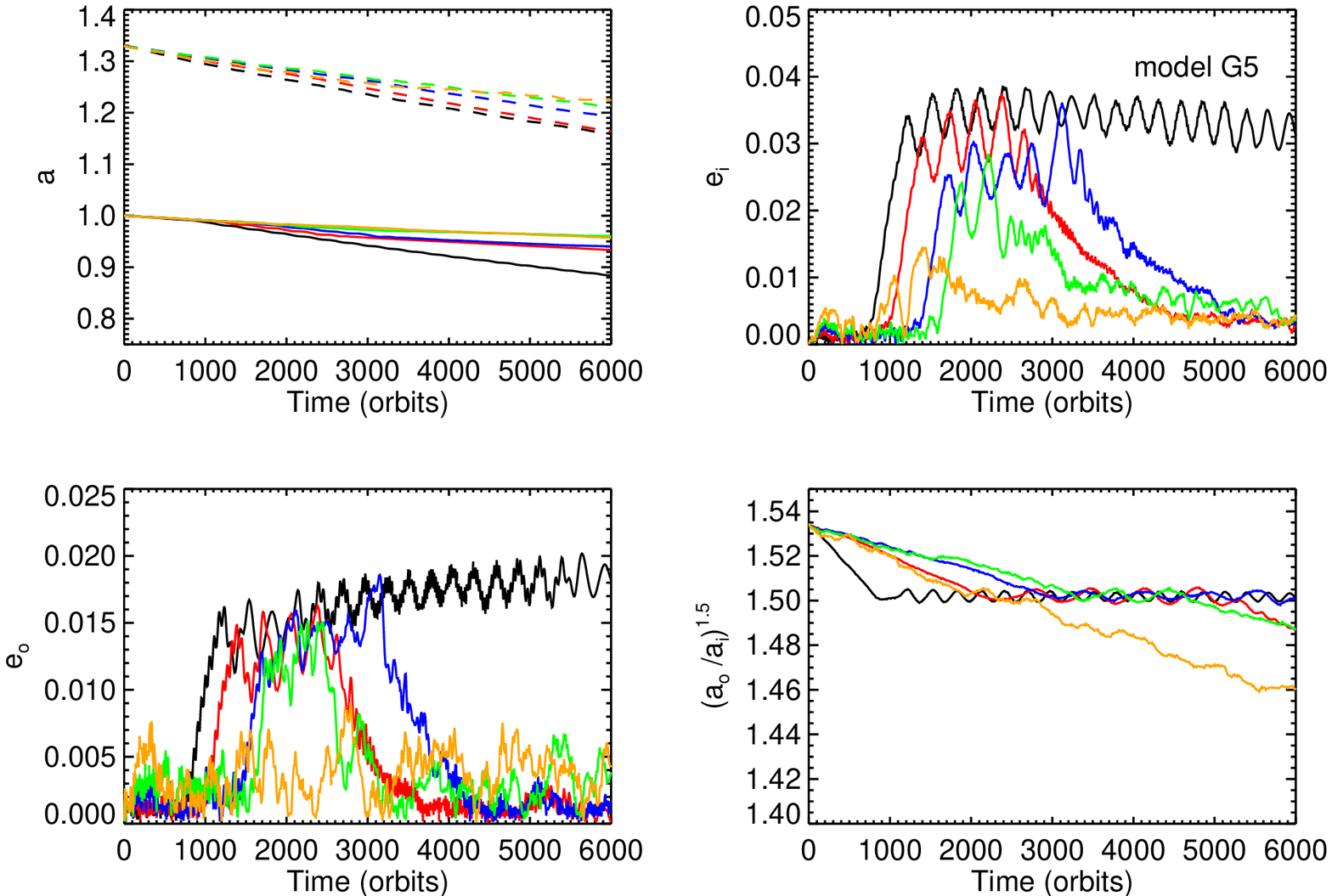}
\caption{{\it Upper left (first) panel}: time evolution of planet
semi-major axes for model G5 and for the different values of $\gamma$ we considered namely for
$\gamma=0$ (black line), $\gamma=6\times 10^{-5}$ (red line), $\gamma=1.3\times 10^{-4}$ (blue),
$\gamma=1.9\times 10^{-4}$ (green) and $\gamma=3\times 10^{-4}$ (orange). {\it Upper right (second)
panel}: time evolution of the inner planet ecentricity. {\it Third panel:} time evolution of the
outer planet eccentricity. {\it Fourth panel:} time evolution of the period ratio
$(a_o/a_i)^{1.5}$. Simulations were performed with GENESIS.}
\label{run5}
\end{figure*}

\begin{figure}
\centering
\includegraphics[width=\columnwidth]{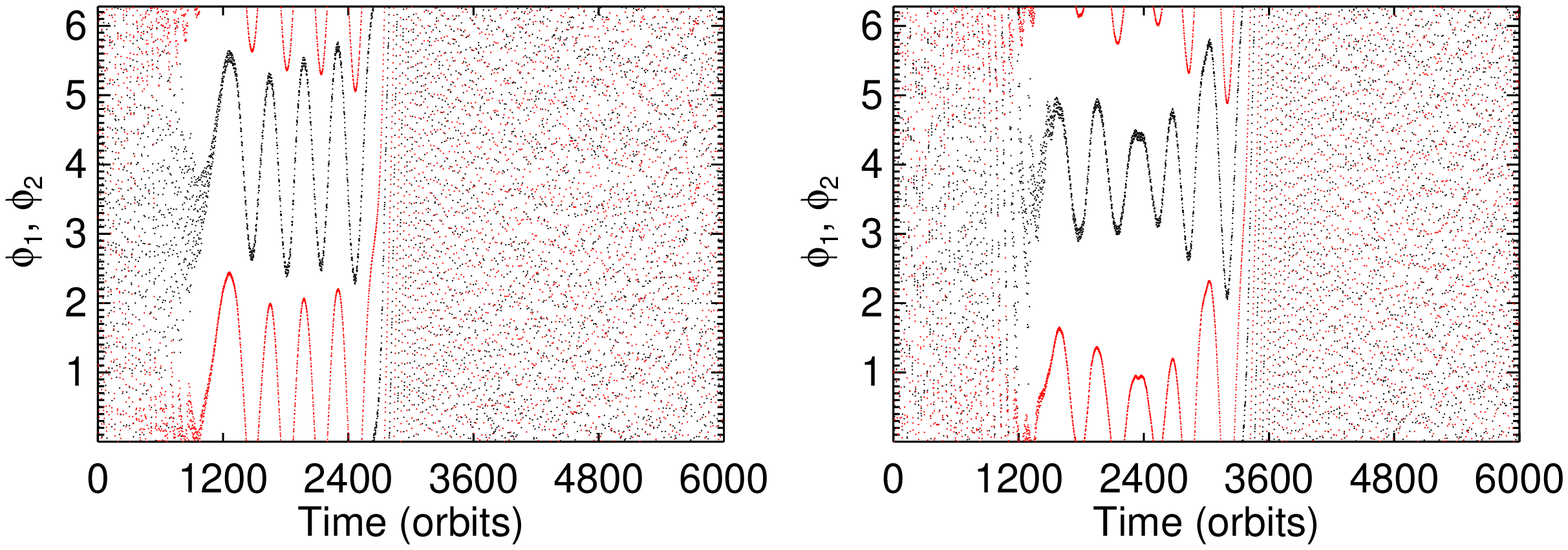}
\caption{Time evolution of the
resonant angles $\phi_1=3\lambda_o-2\lambda_i-\omega_i$ (black) and
$\phi_2=3\lambda_o-2\lambda_i-\omega_o$ (red) for model G1 with $\gamma=6\times 10^{-5}$ (left panel) and
$\gamma=1.3\times 10^{-4}$ (right panel).}
\label{run5-angles}
\end{figure}

\begin{figure}
\centering
\includegraphics[width=\columnwidth]{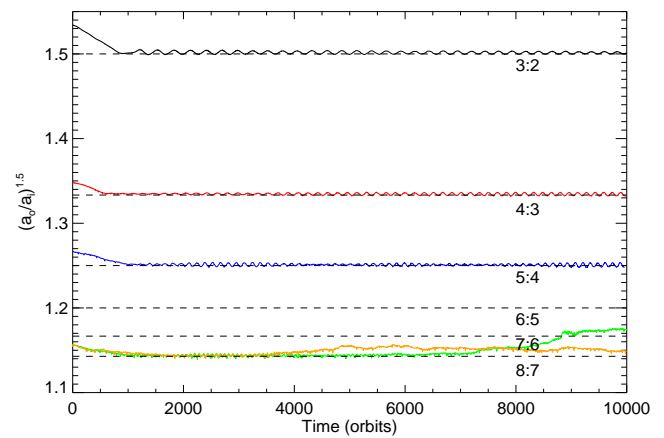}
\caption{Period ratio between the two planets
for model G5 with $\gamma=0$ (black line), $\gamma=6\times 10^{-5}$ (red), $\gamma=1.3\times 10^{-4}$ (blue),
$\gamma=1.9\times 10^{-4}$ (green) and $\gamma=3\times 10^{-4}$ (orange). Simulations were 
performed with GENESIS.}
\label{per-run5}
\end{figure}

For systems with $q=1/2$, the results of the simulations indicate that the 3:2 resonance can be maintained only 
in cases where the disk is close to being inviscid.
Fig. \ref{run5} shows the results for model G5 in which $m_i=1.6$ $M_\oplus$ and
$m_o=3.3$ $M_\oplus$ and for a single realization of the different values of $\gamma$ we considered. 
Moving from left to right and from 
top to bottom the panels display the time evolution of the planets' semi-major axes, eccentricity of the inner 
planet, eccentricity of the outer one, and period ratio. The evolution of semi-major axes shows strong 
similarities with that of model G1, with the migration rates of the planets observed
to decrease with increasing the value for $\gamma$, as discussed in Sect. \ref{sec:orbit}. \\ 
Because of stochastic density fluctuations, the planets eccentricities are highly variables 
quantities and a clear trend of higher eccentricities for higher values of $\gamma$ is again observed at 
the beginning of the simulations. For $\gamma=0$, the time evolution of the period ratio 
shows that trapping in the 3:2 resonance occurs at 
$t\sim 10^3$ orbits. This resonant interaction causes eccentricity growth of both 
planets with the eccentricities of the inner and outer planets reaching peak values of 
$e_i\sim 0.035$ and $e_o\sim 0.02$ respectively, although convergence is not fully established at the end of 
the simulation. Comparing Figs \ref{run1} and \ref{run5}, we see that the period ratio 
oscillates with greater amplitude in model G5,
indicating thereby that resonant locking is weaker. This arises because, compared with model G1, the relative 
migration rate is higher for that model. Here, it is worthwhile to notice that over timescales longer that those
covered by the simulations, it is not clear whether or not the planets will remain bound in the 3:2 resonance
since for disparate planet masses as is the case for model G5, we expect the dynamics of the system to 
be close to the chaotic regime even for $\gamma=0$ (Papaloizou \& Szuszkiewicz 2005).\\ 
For turbulent runs, we find that the planets become temporarily trapped in the 3:2 resonance but in each case, 
the final outcome appears to be disruption of that resonance. Not surprisingly, the survival time of the
3:2 resonance tends to increase with decreasing the value for $\gamma$. 
For these realizations, the lifetimes of the resonance are estimated to be $\sim 1500$ orbits for $\gamma=6\times 10^{-5}$, 
$\sim 2\times 10^3$ orbits for $\gamma=1.3\times 10^{-4}$, $\sim 10^3$ orbits
for $\gamma=1.9\times 10^{-4}$. The slightly longer lifetime of the resonance obtained in the run 
with $\gamma=1.3\times 10^{-4}$ arises because of the stochastic nature of the problem. \\ 
In Fig. \ref{run5-angles} is displayed for two runs with $\gamma=6\times 10^{-5}$ and 
$\gamma=1.3\times 10^{-4}$ the evolution 
of the resonant angles associated with the 3:2 resonance.  In comparison with models in which $q=1$, there is 
a clear tendancy for these to librate with higher amplitude.Therefore, for models with disparate planet masses, 
the origin of the disruption of 
the 3:2 resonance in turbulent runs is likely to be related to the 
combined effect of diffusion of the resonant 
angles plus high libration amplitudes due to the resonance being weaker. \\
For $\gamma \ge 6\times 10^{-5}$ , we can not rule out the possibility that the planets
would become locked in stronger $p+1$:$p$ resonances with $p\ge 3$. To investigate this issue in more details, 
we have performed an additional set of simulations in which, for each value of $\gamma$ we 
considered,  the outer  planet was initially located just outside the 4:3 resonance with the inner one. 
In cases where the 4:3 resonance was found to be unstable, we performed an additional run but with an
initial separation between the two planets slightly larger than that corresponding to the 5:4
resonance. If the 5:4 resonance is not maintained, we repeat the procedure described above until 
a stable $p+1$:$p$ commensurability is eventually formed. Because performing several realizations for 
each value of $\gamma$ would require a large suite of simulations, we have considered here only 
one single realization for each value of $\gamma$.  \\
Fig. \ref{per-run5} illustrates how the established resonance depends on the value for 
the turbulent forcing. Not surprisingly, a clear trend for forming stronger $p+1$:$p$ 
resonances with increasing $\gamma$ is observed. For $\gamma=6\times 10^{-5}$, the system enters 
in a stable 4:3 resonance while for $\gamma=1.3\times 10^{-4}$, the planets become rather locked in the
5:4 resonance. For the runs with $\gamma=1.9\times 10^{-4}$ and 
$\gamma=3\times 10^{-4}$ however,  the planets become temporarily trapped in the 8:7 resonance but in each case 
this commensurability is subsequently lost  with the planets undergoing 
divergent migration. In that case, it is interesting to note that the system is close to 
the stability limit since for planets in the Earth mass range as is the case here, we expect resonance overlap to 
occur for $p\ge 8$ (Papaloizou \& Szuszkiewicz 2005). Therefore, we can reasonably suggest that for 
such values of $\gamma$, super-Earths with mass ratio $q=m_i/m_o<1/2$ 
may not be able to become trapped in a stable mean motion resonance and may eventually suffer close encouters.

\section{Discussion and conclusion}

In this paper we have presented the results of hydrodynamic simulations aimed
at studying the evolution of a system composed of two planets in the Earth mass
range and embedded in a turbulent protoplanetary disk. We employed the turbulence 
model of Laughlin et al. (2004)  and modified by Baruteau \& Lin (2010) in which a turbulent potential corresponding to the
superposition of multiple wave-like modes is applied to the disk. We focused on a scenario
in which the outermost planet is initially located just outside the 3:2 resonance and 
investigated how the evolution depends on both the planet mass ratio $q$ and the value for 
the turbulent forcing parameter $\gamma$. \\
The results of the simulations indicate that for systems with equal mass planets, a 3:2 
resonance can be maintained in presence of weak turbulence.  For instance, in the case 
 of two planets with equal mass $3.3$ $M_\oplus$, we find that the 3:2 resonance is stable in runs 
with $\gamma \le 1.9\times 10^{-4}$, which corresponds to values for the effective viscous stress 
parameter of $\alpha \lesssim 2\times 10^{-3}$. Such a value was found to compare 
fairly well with that resulting from both analytical estimations and preliminary N-body 
runs. 
For systems with planet mass ratios $q\le 1/2$ however, it appears that a 3:2 resonance can remain 
stable only provided that the disk is close  to being inviscid. In turbulent disks however, 
the outcome depends strongly on the value for $\gamma$:\\
i) For $ 6\times 10^{-5}\le\gamma\le 1.3\times 10^{-4}$ 
(equivalent to $ 2\times 10^{-4}\lesssim\alpha\lesssim 10^{-3}$) , 
the planets tend to become locked in stronger $p+1$:$p$ resonances, with $p$ increasing 
as the value for $\gamma$ increases.\\
ii) In the case where $\gamma\ge 1.9\times 10^{-4}$ (equivalent to 
$\alpha\ge 2\times 10^{-3}$), we 
find that the planets can become temporarily trapped in a 8:7 commensurability, but this
resonance is 
disrupted at later times and no stable resonance is formed.\\
Given that the volume averaged stress parameter deduced from MHD simulations is 
typically $\alpha\sim  5\times 10^{-3}$ (Papaloizou \& Nelson 2003; Nelson 2005), these 
results suggest that mean motion resonances between planets in the Earth mass range 
are likely to be disrupted in the 
active zones of protoplanetary disks. For relatively low levels of turbulence however, 
as is the case for a dead-zone (Gammie 1996), a resonance can be maintained for 
moderate values of the planet mass ratio. \\
Such a scenario is broadly consistent with the preliminary analysis of $\sim 170$ 
multi-planetary systems candidates recently detected by 
Kepler (Lissauer et al. 2011) and which suggests that only a few of the 
observed adjacent pairings are either in or near a MMR. However, examination of the 
slope of the cumulative distribution of period ratios (Fig. 7 of Lissauer et al. 2011) 
also reveals an excess of planets with period ratios corresponding to the 2:1 or 
3:2 commensurabilities. In that case, it appears that the neighboring planet 
candidates have masses within $20\%$ of each other. This clearly supports our 
findings that in disks with moderate levels of turbulence, MMRS are stable provided the
mass ratio between the neighboring planets is close to unity. 
 \\
Since turbulence has a significant impact on the capture of two planets in the Earth mass 
range, it will be of interest to examine this issue using three-dimensional MHD 
simulations, which eventually include the presence of a dead-zone. We will address this 
issue in a future paper.  

\end{document}